\documentclass{article}

\usepackage{arxiv}

\usepackage[utf8]{inputenc} 
\usepackage[T1]{fontenc}    
\usepackage{hyperref}       
\usepackage{url}            
\usepackage{booktabs}       
\usepackage{amsfonts}       
\usepackage{nicefrac}       
\usepackage{microtype}      
\usepackage{lipsum}		
\usepackage{natbib}
\usepackage{braket}
\usepackage{graphicx}
\usepackage{amsmath}
\usepackage{amssymb}
\usepackage{amsfonts}
\usepackage{braket}
\usepackage{float}
\usepackage[table,xcdraw]{xcolor}
\usepackage{cite}
\usepackage{subfig}

\usepackage{setspace}
\usepackage{multirow}

\usepackage{algorithmic}
\usepackage{textcomp}
\usepackage{xcolor}
\usepackage{rotating}
\usepackage{booktabs}

\title{Search for Multiple Adjacent Marked Vertices on the Hypercube by a Quantum Walk with Partial Phase Inversion}

\author{ \href{https://orcid.org/0000-0002-6527-7065}{\includegraphics[scale=0.06]{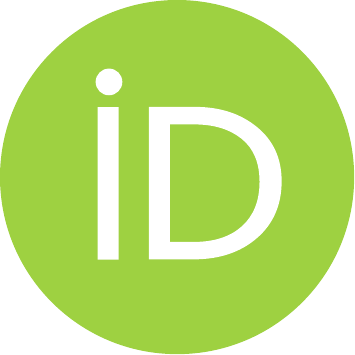}\hspace{1mm}Luciano S. de Souza}\thanks{R. Dom Manuel de Medeiros, s/n, Dois Irmãos -- Recife, Pernambuco -- Brasil} \\
	Departamento de Estat\'{i}stica e Inform\'{a}tica\\
	Universidade Federal Rural de Pernambuco\\
	Recife, Brasil \\
	\texttt{luciano.serafim@ufrpe.br} \\
	\And
	\href{https://orcid.org/0000-0002-2672-7801}{\includegraphics[scale=0.06]{orcid.pdf}\hspace{1mm}Jonathan H. A. de Carvalho} \\
	Centro de Inform\'{a}tica\\
	Universidade Federal de Pernambuco\\
	Recife, Brasil \\
	\texttt{jhac@cin.ufpe.br} \\
 \And
	\href{https://orcid.org/0000-0002-9544-7774}{\includegraphics[scale=0.06]{orcid.pdf}\hspace{1mm}Henrique C. T. Santos} \\
 Instituto Federal de Educação, Ciência e\\ Tecnologia de Pernambuco\\
	Recife, Brasil \\
	\texttt{henrique.santos@recife.ifpe.edu.br}
	\And
	\href{https://orcid.org/0000-0002-2131-9825}{\includegraphics[scale=0.06]{orcid.pdf}\hspace{1mm}Tiago A. E. Ferreira} \\
	Departamento de Estat\'{i}stica e Inform\'{a}tica\\
	Universidade Federal Rural de Pernambuco\\
	Recife, Brasil \\
	\texttt{tiago.espinola@ufrpe.br} \\
}

\renewcommand{\shorttitle}{Search for Multiple Adjacent Marked Vertices on the Hypercube by a QW with PPI}

\hypersetup{
pdftitle={A template for the arxiv style},
pdfsubject={q-bio.NC, q-bio.QM},
pdfauthor={David S.~Hippocampus, Elias D.~Striatum},
pdfkeywords={First keyword, Second keyword, More},
}

\begin{document}
\maketitle

\begin{abstract}
Quantum walks provide an efficient tool for the construction of new quantum search algorithms. In this paper, we analyze the application of the Multiself-loop Lackadaisical Quantum Walk on the hypercube that uses partial phase inversion of the target state to search for multiple adjacent marked vertices. We consider two scenarios and one of them evaluates the influence of the relative position of non-adjacent marked vertices on the search results. The use of self-loops and the composition of their weights are an essential part of the construction process of new quantum search algorithms based on lackadaisical quantum walks, however, other aspects have been considered, such as, for example, the type of marked vertices. It is known that part of the energy of a quantum system is retained in states adjacent to the target state. This behavior causes the amplification of these states where the sum of probability amplitudes reaches values equivalent to those of the target state, reducing their chances of being observed. Here we show experimentally that with the use of partial phase inversion of the target state, it is possible to amplify its probability amplitudes even in scenarios with adjacent marked vertices reaching maximum success probabilities with values close to $1$. We also show that the relative position of the non-adjacent marked vertices did not significantly influence the results. The lackadaisical quantum walk generalization in using multiple self-loops to only a single self-loop and the ideal composition of a weight value was sufficient to obtain advances to quantum search algorithms based on quantum walks. However, the results presented here show that many other aspects need to be taken into account for the construction of new quantum algorithms. It was possible to add gains in the maximum probabilities of success compared to other results found in the literature. In one of the most significant cases, the probability of success increased from $p \approx 0.38$ to $p > 0.99$. Therefore, the use of partial phase inversion of target states brings new contributions to the development of new quantum search algorithms based on quantum walks and the use of multiple self-loops.
\end{abstract}

\keywords{Quantum Computing \and Quantum Walks \and Quantum Search Algorithm \and Lackadaisical Quantum Walk \and Multiple Self-loops \and Adjacent Marked Vertices \and Partial Phase Inversion}
 
\section{Introduction}

Many advances have been achieved since the publication of the article by \citet{aharonov1993quantum}, which is considered the first in quantum walks. One of the first quantum search algorithms based on quantum random walks was designed by \citet{shenvi2003quantum}, which defined the quantum walks as one of the most promising resources and an intuitive framework for building new quantum algorithms. Many other works on quantum walks have been developed since this moment \citep{ambainis2004coins,potovcek2009optimized,hein2009quantum,ambainis2012search}.

Amongst many proposed works in quantum walks, \citet{wong2015grover} developed a quantum search algorithm called lackadaisical quantum walks - LQW, an analog of classical lazy random walks  in which the quantum walker has a chance to stay at the current vertex position by introducing $m$ self-loops of integer weight $l$ at each vertex of the complete graph. This proposal was altered by \citet{wong2017coined} where the $m$ self-loops were generalized to one self-loop of non-integer weight. In turn, \citet{desouza2023multiselfloop} proposed a new quantum search algorithm based on the LQW called Multiself-Loop Lackadaisical Quantum Walk - MSLQW, which uses $m$ self-lops in each vertex on the hypercube with weight value $l = l'\cdot m$, and the partial phase inversion of the target state to research multiple marked vertices. The weight value $l' \in \mathbb{R}$ and $m \in \mathbb{Z}$.

However, some other studies indicate that the type of marked vertices influences the results of quantum search algorithms, in particular, the adjacent marked vertices.
According to \citet{potovcek2009optimized}, the final state of the algorithm designed by \citet{shenvi2003quantum} is mainly composed of the marked state but also has small contributions from its nearest neighbors, i.e., part of the probability amplitude is retained in adjacent vertices. Another behavior of quantum walks on the hypercube referring to adjacent marked vertices is the formation of stationary states \citep{nahimovs2019adjacent}. \citet{souza2021lackadaisical} experimentally showed that adjacent marked vertices interfere with the results of the search for multiple marked vertices. Although they have proposed a new ideal value of weight $ l = (d/N) \cdot k$, when there are adjacent marked vertices in the set of solutions occurs a decrease in the maximum probability of success.

Therefore, this work objective is to apply MSLQW-PPI to research multiple marked vertices in two scenarios. The first scenario analyzes research by multiple adjacent and no-adjacent vertices to verify that the relative position of non-adjacent vertices interferes with the search results. The second scenario analyzes research by multiple adjacent vertices. Based on the methodology used by \citet{desouza2023multiselfloop}, the results presented in this work are promising. Comparing the results of this work with the results obtained by \citet{souza2021lackadaisical} there was a gain in the maximum probability of success to values close to $1$. Before some of the success probabilities reached only $p \approx 0.59$ and $ p \approx 0.38$, the first and second scenarios respectively.

In their proposal, \citet{desouza2023multiselfloop} used the search for non-adjacent marked vertices with the phases of $1 \leqslant s < m$ inverted self-loops and $1 \leqslant m \leqslant 30$. The results indicated that, by inverting the phase of only one self-loop, it is possible to achieve maximum probabilities of success close to $1$. Based on the results of \citet{desouza2023multiselfloop}  we applied the particular case of MSLQW-PPI with $s=1$ inverted self-loop. Compared to the results obtained in this work, the maximum success probabilities remain close to $1$. The coefficient of variation was also used to evaluate the dispersion around the average relative position of the non-adjacent marked vertices. The coefficient of variation was also used to evaluate the dispersion around the average maximum probability according to the relative position of the non-adjacent marked vertices. The results indicate that the variation around the maximum mean probability of success is not significant. These results are important because they show that the partial inversion of the target state based on the use of multiple self-loops provides a new perspective of advances in the development of new quantum search algorithms.


This paper is organized as follows. In Section \ref{sec:mslqw-on-the-hypercube} we present some concepts about Multiself-loop Lackadaisical Quantum Walks in the hypercube. Section \ref{sec:experiments-setup} the experiments are defined. Section \ref{sec:results-and-discussion} presents the results and discussion. Finally, in Section \ref{sec:conclusions} are the conclusions.

\section{Multi-self-loop lackadaisical quantum walk on the hypercube}
\label{sec:mslqw-on-the-hypercube}

The lackadaisical quantum walk is the quantum counterpart of the classical lazy random walk. This quantum algorithm was proposed by \citet{wong2015grover} and is obtained by adding a self-loop to each vertex of the graph. Then, the Hilbert space associated with the lackadaisical quantum walk in the hypercube is

\begin{equation*}
    \mathcal{H} = \mathcal{H}^{n+1} \otimes \mathcal{H}^{2^{n}}
\end{equation*}
where $\mathcal{H}^{n+1}$ is the Hilbert space associated with the quantum coin space, and $\mathcal{H}^{2^{n}}$ is the Hilbert space associated with nodes in the hypercube. According to \citet{hoyer2020analysis}, in a $n$-regular graph by adding a self-loop of weight $l$ to each vertex, the coined Hilbert space becomes

\[\mathcal{H}^{n+1} = \{\ket{e_{0}}, \ket{e_{1}}, \dots, \ket{e_{n-1}}, \ket{\circlearrowleft}\}.\]
where $e_{i}$ is a binary string of $n$ bits with $1$ in the $i$-th position \citep{kempe2002quantum,shenvi2003quantum}, and $\ket{\circlearrowleft}$ is the self-loop. Weighted self-loop accounting is done by modifying Grover's coin as follows

\begin{equation}
\label{eq:grovers-coin-self-loop}
    C = 2\ket{s^{C}}\bra{s^{C}} - I_{(n + 1)}
\end{equation}
where 

\begin{equation}
\label{eq:s-c-add-self-loop}
    \ket{s^{C}} = \frac{1}{\sqrt{n + l}} \left ( \sqrt{l}\ket{\circlearrowleft} + \sum_{i=0}^{n-1}\ket{i} \right ).
\end{equation}
The Lackadaisical Quantum Walk system in the hypercube starts as follows

\begin{equation}
\label{eq:initial-system-lqw}
    \ket{\Psi(0)} = \frac{1}{\sqrt{N}} \sum_{\vec{x}}\ket{\vec{x}} \otimes \ket{s^{C}}.
\end{equation}
Substituting Equation \ref{eq:s-c-add-self-loop} into Equation \ref{eq:initial-system-lqw} we obtain the initial state described in Equation \ref{eq:initial-state-lqw}.

\begin{equation}
    \label{eq:initial-state-lqw}
    \ket{\Psi(0)} = \frac{\sqrt{l}}{\sqrt{N} \times \sqrt{n+l}} \sum_{\Vec{x}} \ket{\Vec{x},\circlearrowleft} + \frac{1}{\sqrt{N} \times \sqrt{n+l}} \sum_{\Vec{x}} \sum_{i=0}^{n-1} \ket{\Vec{x},i}
\end{equation}

The Multiself-loop lackadaisical quantum walk was proposed by \citet{desouza2023multiselfloop}. This quantum algorithm is obtained by adding $m$ self-loops at each vertex of the hypercube and a partial phase inversion of the target state is applied. The Hilbert space associated with the lackadaisical quantum walk in the hypercube is

\begin{equation*}
    \mathcal{H} = \mathcal{H}^{n+m} \otimes \mathcal{H}^{2^{n}}.
\end{equation*}
Then, the Hilbert space associated with the coin space becomes

\[\mathcal{H}^{n+m} = \{\ket{e_{0}}, \ket{e_{1}}, \dots, \ket{e_{n-1}}, \ket{\circlearrowleft_{0}}, \ket{\circlearrowleft_{1}}, \dots, \ket{\circlearrowleft_{m-1}}\}.\]
To account for the weighted auto-loop, a modification is made to the Grover coin described in Equation~\ref{eq:grovers-coin-self-loop} as follows

\begin{equation}
\label{eq:grovers-coin-m-self-loop}
    C = 2\ket{s^{C}}\bra{s^{C}} - I_{(n + m)}
\end{equation}
where

\begin{equation}
\label{eq:add-mult-self-loop}
    \ket{s^{C}} = \frac{1}{\sqrt{n + l}} \left (\sqrt{l'} \sum_{j=0}^{m-1} \ket{\circlearrowleft_{i}} + \sum_{i=0}^{n-1}\ket{i} \right )
\end{equation}
and $l = l' \cdot m$. The Multiself-loop lackadaisical quantum walk system on the hypercube is also started according to Equation \ref{eq:initial-system-lqw}. Substituting Equation \ref{eq:add-mult-self-loop} into Equation \ref{eq:initial-system-lqw}, we obtain the initial state described in Equation \ref{eq:initial-state-mslqw}.

\begin{equation}
\label{eq:initial-state-mslqw}
\ket{\Psi(t = 0)} = \frac{\sqrt{l'}}{\sqrt{n+l} \times \sqrt{N}} \sum_{j=0}^{m-1}\sum_{\vec{x}} \ket{\circlearrowleft_{j},\vec{x}} + \frac{1}{\sqrt{n+l} \times \sqrt{N}} \sum_{i=0}^{n-1}\sum_{\vec{x}}\ket{i,\vec{x}}
\end{equation}
The proposed modification of the oracle described in Equation \ref{eq:oracle-mslqw} makes it possible to identify the components of the target state.

\begin{align}
\label{eq:oracle-mslqw}
   Q = I_{(n+\text{m})\cdot N}  - 2 \sum_{\omega}\sum_{\epsilon = 1}^{n} \ket{\epsilon,\omega}\bra{\epsilon,\omega}
   - 2\sum_{\omega}\sum_{\tau} \ket{\circlearrowleft _{\tau},\omega}\bra{\circlearrowleft _{\tau},\omega}
\end{align}
where $\ket{\omega}$ represents the marked vertex, $\epsilon$ represents an edge that is not a self-loop, and $\circlearrowleft_{\tau}$ is the self-loop that will have its phase inverted.
Consider an arbitrary state which denotes the superposition of all edges

\begin{align}
\begin{split}
\label{eq:arbitrary-state}
\ket{\textbf{x}} &= \ket{\vec{x},\circlearrowleft_{0}}+\ket{\vec{x},\circlearrowleft_{1}}+\ket{\vec{x},\circlearrowleft_{2}}+ \dots + \ket{\vec{x}, \circlearrowleft_{m-1}}+ \ket{\vec{x},0}+ \ket{\vec{x},1}+ \ket{\vec{x},2}+\dots+ \ket{\vec{x},n-1}.
\end{split}
\end{align}
Consider the states $\ket{\circlearrowleft _{\tau}}$ as the target self-loops and $s=1$. Applying the phase inversion operator, represented by Equation \ref{eq:phase-inversion-operator}.

\begin{equation}
\label{eq:phase-inversion-operator}
    Q = I_{(n+\text{m})\cdot N} - 2\sum_{\omega}\sum_{\epsilon = 0}^{n-1}\ket{\epsilon,\omega}\bra{\epsilon,\omega}\\
   -2\sum_{\omega}\sum_{\tau}\ket{\circlearrowleft_{\tau=0},\omega}\bra{\circlearrowleft_{\tau=0},\omega}
\end{equation}
where $I_{n+m}$ is the identity operator of dimension $n+m$ and $\ket{\circlearrowleft _{\tau=0}}$ as the target self-loop, we have

\begin{align}
\begin{split}
\label{eq:arbitrary-state}
\ket{\vec{\textbf{x}}} &= -\ket{\vec{x},\circlearrowleft_{0}}+\ket{\vec{x},\circlearrowleft_{1}}+\ket{\vec{x},\circlearrowleft_{2}}+ \dots + \ket{\vec{x}, \circlearrowleft_{m-1}}- \ket{\vec{x},0} - \ket{\vec{x},1} - \ket{\vec{x},2} - \dots - \ket{\vec{x},n-1}.
\end{split}
\end{align}

\section{Experiment setup}
\label{sec:experiments-setup}

According to \citet{shenvi2003quantum,potovcek2009optimized}, part of the probability amplitude of a quantum walk on the hypercube is retained at vertices adjacent to a marked vertex, and if two marked vertices on the hypercube are adjacent stationary states are formed \citep{nahimovs2019lackadaisical}. \citet{souza2021lackadaisical} showed that adjacent marked vertices interfere with the search performance. Therefore, the experiments performed in this work are divided into the following two scenarios. In the first scenario, we consider both adjacent and non-adjacent marked vertices. In the second scenario, we consider only the adjacent marked vertices.

\subsection{Definition of marked vertex samples}
\label{sec:definition-of--marked-vertex-samples}

According to the definitions of the hypercube, two vertices are adjacent if the Hamming distance between them is $1$. Non-adjacent vertices are those that have a Hamming distance of at least $2$ from any other vertex. In this way, we define the set of marked vertices to execute the simulations. The set of marked vertices is divided into $M_{k,\gamma}$ groups of samples with $k$ vertices and $j$ samples.


The first set is formed by the vertices that are both adjacent and non-adjacent. This set is divided into twelve groups of one hundred samples. For each sample of $k$ adjacent vertices, other $k-1$ non-adjacent vertices are also marked, and thirty MSLQW-PPI are performed. Therefore, thirty-six hundred simulations are performed. Every hundred simulations we preserve the same $k$ adjacent marked vertices and vary the locations of the $k-1$ non-adjacent marked vertices, for example, if $k = 3$ we have two adjacent marked vertices and one non-adjacent marked vertex,

\[M_{3,100} = [\{0, 1, 1128\}_{1},\{0, 1, 2950\}_{2},\dots,\{0, 1, 1470\}_{ 100}] \]
Every one hundred new simulations, $k$ new adjacent vertices and $(k-1)\cdot 100$ new non-adjacent vertices are marked and added to the group until $k + (k-1) = 5, 7, \dots, 25$.


The second set is formed by the adjacent vertices. This set is divided into twelve groups of one sample that contain between $2$ and $13$ marked vertices. To search for adjacent vertices, twelve simulations are performed. Initially, we have two marked vertices and at each new simulation a new vertex is marked and added to the new group as follows

\[M_{2,1} = \{0,1\}, M_{3,1} =  \{0,1,2\},\dots,M_{13,1} = \{0,1,2,\dots,1024,2048\}\]
until all adjacent vertices are marked.

The samples have $k$ distinct vertices, i.e., without replacement. The simulations performed in the set of the first scenario were necessary so that we could obtain the average behavior based on the relative position of the non-adjacent marked vertices and verify their influence on the results. In each simulation, thirty MSLQW-PPI are performed. The stop condition for a simulation occurs after each of the thirty walks obtains the maximum value of the probability amplitude. In each quantum walk, a number $m$ of self-loops per vertex was defined, which varies between $1$ and $30$. The weight $l$ is distributed by dividing its value between $m$ equal parts.

\subsection{Hardware and software setup for the simulations}
\label{sec:hardware-software-setup}

The simulations were performed using the Parallel Experiment for Sequential Code - PESC~\citep{santos2023pesc}, to perform computational simulations distributed over a network. The platform provides a web interface for configuring simulation requests and manages the status and lifecycle of the request. The use of the platform simplified the simulations execution process, which was important to support the collect the study data. The tool is being developed for the instrumentation and optimization of the research group's computational experiments. The programming language used to write the algorithms was Python 3.7. All machines that were used in the simulations utilize the operational system,  Ubuntu 18.04.6 LTS (Bionic Beaver), and have an HD of 500 GiB. Table~\ref{tab:machine-settings} shows the machines' settings.

\begin{table}[h]
\centering
\caption{Machine hardware configuration.}
\caption*{\textbf{Adapted from:} \citep{desouza2023multiselfloop}.}
\begin{tabular}{crl}
\toprule
\textbf{Machines} & \multicolumn{1}{l}{\textbf{System RAM}} & \textbf{System Processor}\\ \midrule
1  & 8  GiB & Intel(R) Core(TM) i7-2600 CPU @ 3.40GHz  \\
1  & 16 GiB & Intel(R) Core(TM) i7-8700 CPU @ 3.20GHz  \\
2  & 32 GiB & Intel(R) Core(TM) i7-2600K CPU @ 3.40GHz \\
2  & 32 GiB & Intel(R) Core(TM) i7-6700 CPU @ 3.40GHz  \\ \bottomrule
\end{tabular}%
\label{tab:machine-settings}
\end{table}

\section{Results and discussion}
\label{sec:results-and-discussion}

As previously defined, the experiments are divided into two scenarios according to the type of marked vertices. In the first scenario, we have adjacent and non-adjacent marked vertices. As we analyzed the relative positional of the non-adjacent marked vertices, thirty-six thousand simulations were performed, which are divided into twelve groups with one hundred samples from $k$ vertices, and to each sample was performed thirty quantum walks MSLQW. Then, the variability of the results was also analyzed and is represented in Fig.~\ref{fig:coefficient-variation-neighbors-non-neighbors}. In the second scenario, we only have adjacent marked vertices. Each node has the same number of adjacent vertices as the hypercube's degree number, therefore, twelve simulations were realized, and in each simulation, thirty MSLQW-PPI were also performed. The quantum walks vary according to the number of self-loops from one to thirty. The results are represented in Figures~\ref{fig:probability-distribution-neighbors-non-neighbors} and \ref{fig:probability-distribution-neighbors} respectively. They present the maximum probability of success according to the number of self-loops and marked vertices.

\subsection{Analyzing the search with adjacent and non-adjacent marked vertices}
\label{sec:analyzing-the-search-with-adjacent-and-non-adjacent-marked-vertices}

\begin{figure}[h]
\centering
\subfloat[$l = n/N$]{\includegraphics[width=8cm]{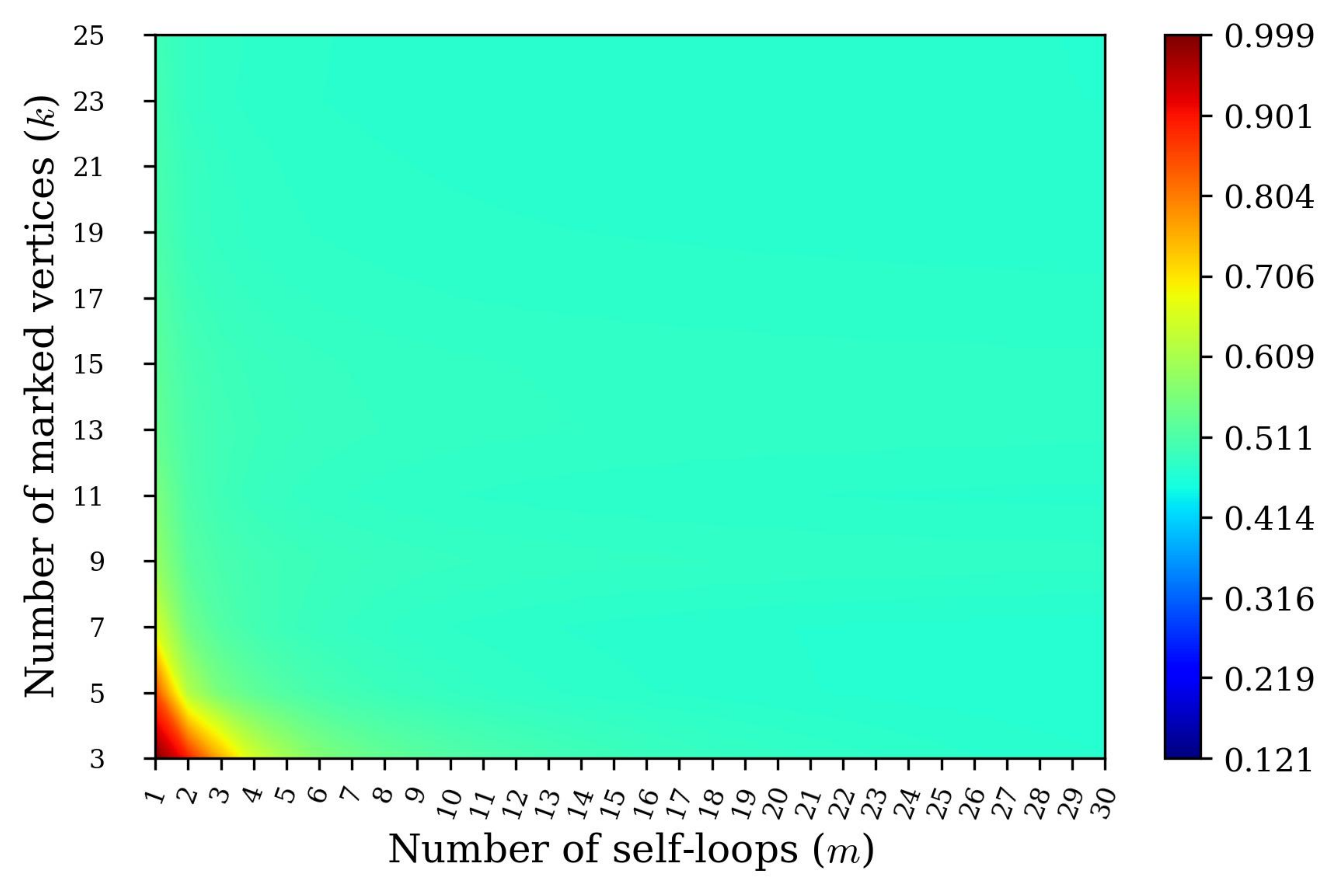}
\label{fig:probability-distribution-neighbors-non-neighbors-a}}
\subfloat[$l = (n/N)\cdot k$]{\includegraphics[width=8cm]{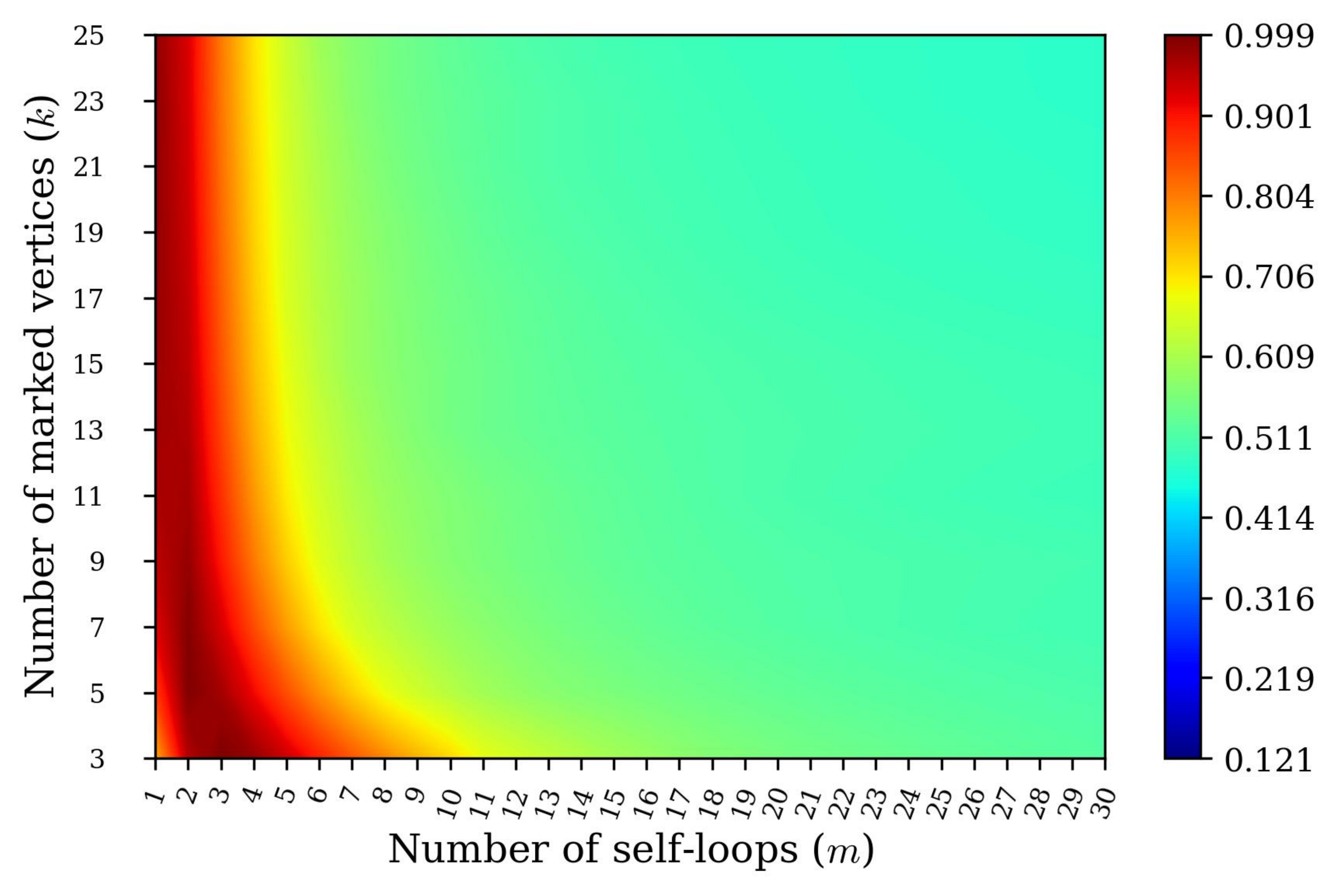}\label{fig:probability-distribution-neighbors-non-neighbors-b}}\\
\subfloat[$l = n^{2}/N$]{\includegraphics[width=8cm]{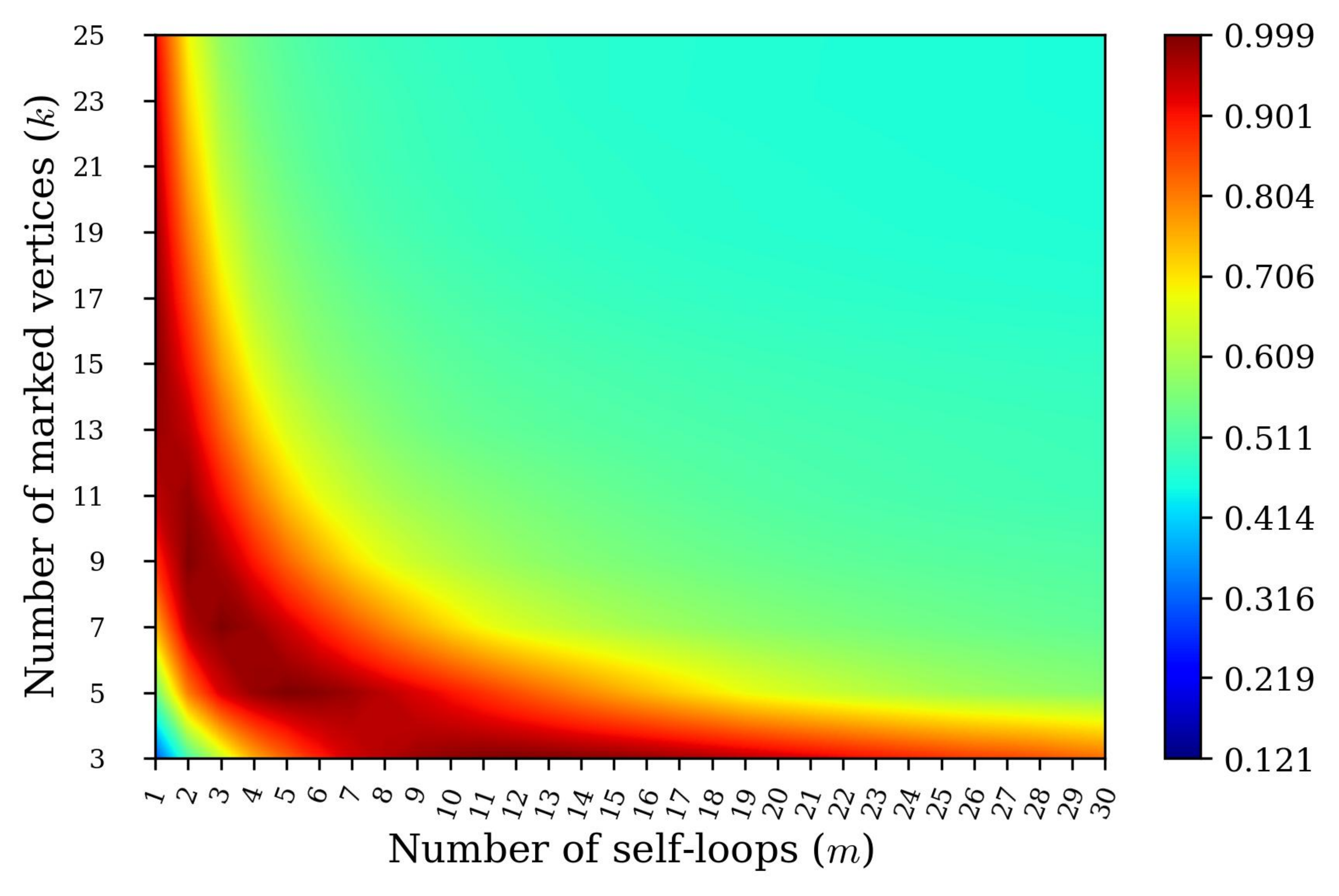}\label{fig:probability-distribution-neighbors-non-neighbors-c}}
\subfloat[$l = (n^{2}/N)\cdot k$]{\includegraphics[width=8cm]{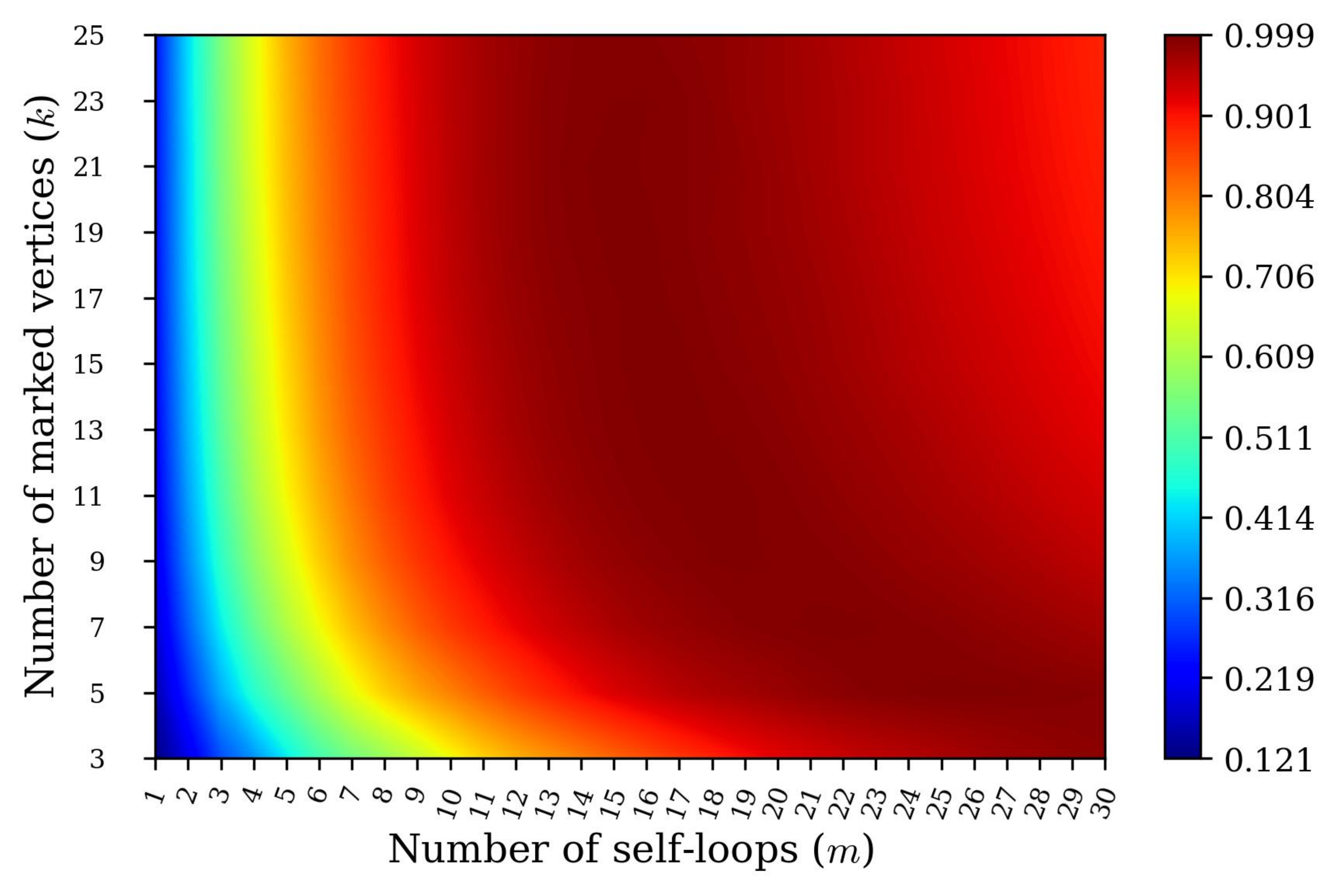}\label{fig:probability-distribution-neighbors-non-neighbors-d}}
\caption{The probability of success of the MSLQW-PPI to search for adjacent and non-adjacent marked vertices with $n = 12$ and $N = 4096$ vertices. \protect\subref{fig:probability-distribution-neighbors-non-neighbors-a} weight value $l = n/N$. \protect\subref{fig:probability-distribution-neighbors-non-neighbors-b} weight value $l = (n/N) \cdot k$. \protect\subref{fig:probability-distribution-neighbors-non-neighbors-c} weight value $l = n^{2}/N$. \protect\subref{fig:probability-distribution-neighbors-non-neighbors-d} weight value $l = (n^{2}/N)\cdot k$.}
\label{fig:probability-distribution-neighbors-non-neighbors}
\end{figure}

Fig.~\ref{fig:probability-distribution-neighbors-non-neighbors-a} shows the probability of success for the weight $l = n/N$. \citet{rhodes2020search} proposed this weight value to search a single vertex while \citet{souza2021lackadaisical} used it to search multiple vertices, however, the results showed that this weight value is not ideal in this case. \citet{desouza2023multiselfloop} used this weight value and applied MSLQW-PPI to search for multiple non-adjacent marked vertices but there was no increase in the maximum probability of success. In this article, the maximum average probability obtained $p = 0.999$ with three marked vertices (two adjacent vertices and one non-adjacent vertex) and a single self-loop, which is a result close to that achieved by \citet{souza2021lackadaisical} of $p = 0.997$. In both cases, as the number of marked vertices increases, the maximum probability of success decreases.

Fig.~\ref{fig:probability-distribution-neighbors-non-neighbors-b} shows the success probability using the weight $l = (n/N)\cdot k$. In cases where there are only non-adjacent marked vertices, for this weight value, only a single self-loop is needed \citep{souza2021lackadaisical,desouza2023multiselfloop}. Although, when there are adjacent marked vertices, it is necessary to increase the number of self-loops to obtain success probabilities close to $1$ as we can see in Table~\ref{tab:weight-and-self-loops-adjacent-and-non-adjacent-n-N-k}.  In some cases, we can observe that there was an improvement in the probability of success compared to the results obtained with the use of only one self-loop. 

Comparing columns A and C of Table \ref{tab:weight-and-self-loops-adjacent-and-non-adjacent-n-N-k}, we can see that from a certain quantity of self-loops, it is possible to obtain more significant probabilities than those achieved with the use of a single self-loop. It is necessary to use the least number of self-loops to obtain these results. Now, comparing columns B and C with at least two self-loops, it is possible to improve the maximum probability of success. However, for this self-loop weight value, as the number of marked vertices increased, only a single self-loop is needed to achieve probabilities of approximately $p \approx 0.98$.

\begin{table}[h]
\centering
\caption{Cases for searching adjacent and non-adjacent marked vertices where more than one self-loop is required to obtain a maximum probability close to $1$ using the weight $l = (n/N)\cdot k$ proposed by \citet{souza2021lackadaisical}. Rows with \text{( - )} in column A mean that the same values are reached in the same rows in column B.}
\label{tab:weight-and-self-loops-adjacent-and-non-adjacent-n-N-k}
\begin{tabular}{cccccccc}
\toprule
& \multicolumn{2}{c}{A} & \multicolumn{2}{c}{B} & \multicolumn{2}{c}{C} \\ \cmidrule(lr){2-3}\cmidrule(lr){4-5}\cmidrule(lr){6-7}
k  &   p   & m &   p   & m &   p   & m \\ \midrule
3  & 0.794 & 8 & 0.999 & 3 & 0.754 & 1 \\
5  & 0.911 & 4 & 0.996 & 2 & 0.863 & 1 \\
7  & 0.931 & 3 & 0.993 & 2 & 0.921 & 1 \\
9  &   -   & - & 0.981 & 2 & 0.948 & 1 \\
11 &   -   & - & 0.970 & 2 & 0.964 & 1 \\
\bottomrule
\end{tabular}
\end{table}

Fig.~\ref{fig:probability-distribution-neighbors-non-neighbors-c} shows the probability of success in the search for adjacent and non-adjacent vertices using the weight value $l = n^ {2}/N$. This weight is proposed by \citet{desouza2023multiselfloop} and is composed of the weight value proposed by \citet{rhodes2020search} to search for one marked vertex plus an exponent in the element that represents the degree of the vertex in the numerator. Compared with the results found by \citet{souza2021lackadaisical} and with the results shown in Fig. \ref{fig:probability-distribution-neighbors-non-neighbors-a}, there was a significant improvement in the maximum probability of success for numbers $k > 3$ marked vertices. In this scenario, the success probabilities depend on the inversely proportional relationship between the number $k$ of marked vertices and the number $m$ of Self-loops. This means that as the number of marked vertices increases, the number of self-loops decreases and the other way around, however, the maximum probability of success continues above $p = 0.97$.

Another analysis of the results for the weight $l = n^{2}/N$ was performed. Two different scenarios are compared and some results are shown in Table~\ref{tab:comp-weight-and-self-loops-adjacent-and-non-adjacent-3c-4c}. The results presented in column A refer to the scenario with only non-adjacent marked vertices obtained by \citet{desouza2023multiselfloop}. In the scenario presented in column B where we have both types of marked vertices for each $k$ adjacent vertices, we have $k - 1$ non-adjacent vertices. Although there are adjacent marked vertices in the sample, the use of multiple self-loops guarantees, in some cases, maximum success probabilities close to $1$.

Note that, in the case where there are only non-adjacent marked vertices, as the number of marked vertices increases the number of self-loops decreases. However, when there are marked adjacent vertices, a larger number of self-loops is needed to maintain the success probability close to the maximum. Comparisons made between different scenarios and the same weights show that the type of marked vertex influences the search result. However, although there are adjacent marked vertices in the sample, partial state inversion guarantees, in some cases, maximum success probabilities close to $1$.

\begin{table}
\centering
\caption{Comparison between the probability of success and number of self-loops for two different scenarios for weight value $l  = n^{2}/N$. Column A represents the results found by \citet{desouza2023multiselfloop} to search for non-adjacent marked vertices and column B to search for adjacent and non-adjacent marked vertices.}
\label{tab:comp-weight-and-self-loops-adjacent-and-non-adjacent-3c-4c}
\begin{tabular}{lcccl}
\toprule
 & \multicolumn{2}{c}{A} & \multicolumn{2}{c}{B} \\ \cmidrule(lr){2-3}\cmidrule(lr){4-5}
k & p & m & p & m \\ \midrule
3 & 0.999 & 4 & 0.999 & 12 \\
5 & 0.990 & 2 & 0.997 & 5 \\
7 & 0.992 & 2 & 0.996 & 3 \\
9 & 0.978 & 1 & 0.996 & 2 \\
11 & 0.996 & 1 & 0.983 & 2 \\ \bottomrule
\end{tabular}
\end{table}

Now, we are going to analyze the case where the scene is the same but the weights are different. Considering the behavior of the probability of success in Figures~\ref{fig:probability-distribution-neighbors-non-neighbors-c} and~\ref{fig:probability-distribution-neighbors-non-neighbors-d}, we can see that not only the type of marked vertices influences the probability of success, but also the weight value. Note that the difference in weight composition, in this case, is the number of marked vertices. We can see that after the increase in the number of self-loops, overall we had a significant improvement in the probability of success. We can better see these results in Table~\ref{tab:weight-and-self-loops-adjacent-and-non-adjacent} with shows the maximum probabilities of success and the number of self-loops according to the number of marked vertices and weight value. Comparing the results described in Table~\ref{tab:weight-and-self-loops-adjacent-and-non-adjacent-n-2-N} and Table~\ref{tab:weight-and-self-loops-adjacent-and-non-adjacent-n-2-N-k}, it is important to realize that the weight composition is very relevant. Although the type of marked vertices can influence the probability of success, with an ideal weight value along with an ideal number of self-loops, it is possible to improve the results. The exception was $k = 3$, where there was a reduction in the probability of success but still close to $1$. The other bold lines show the cases with the more expressive improvements in the probability of success. In general, there was a significant increase in the number of self-loops.

\begin{table}[h]
\centering
\caption{Ideal number of self-loops and maximum probability of success for searching adjacent and non-adjacent marked vertices. (\ref{tab:weight-and-self-loops-adjacent-and-non-adjacent-n-2-N}) weight value $l = n^{2}/N$. (\ref{tab:weight-and-self-loops-adjacent-and-non-adjacent-n-2-N-k}) weight value $l = (n^{2}/N)\cdot k$.}
\label{tab:weight-and-self-loops-adjacent-and-non-adjacent}
\subfloat[]{
\label{tab:weight-and-self-loops-adjacent-and-non-adjacent-n-2-N}
\begin{tabular}{cccccc}
\toprule
k & p & m & k & p & m \\ \cmidrule(lr){1-3}\cmidrule(lr){4-6}
\textbf{3} & \textbf{0.999} & \textbf{12} & 15 & 0.996 & 1 \\
5 & 0.997 & 5 & 17 & 0.991 & 1 \\
7 & 0.996 & 3 & \textbf{19} & \textbf{0.980} & \textbf{1} \\
9 & 0.996 & 2 & \textbf{21} & \textbf{0.960} & \textbf{1} \\
11 & 0.983 & 2 & \textbf{23} & \textbf{0.941} & \textbf{1} \\
13 & 0.983 & 1 & \textbf{25} & \textbf{0.920} & \textbf{1}\\ \bottomrule
\end{tabular}}
\hspace{0.5cm}
\centering
\subfloat[]{
\label{tab:weight-and-self-loops-adjacent-and-non-adjacent-n-2-N-k}
\begin{tabular}{cccccc}
\toprule
k & p & m & k & p & m \\ \cmidrule(lr){1-3}\cmidrule(lr){4-6}
\textbf{3} & \textbf{0.988} & \textbf{30} & 15 & 0.997 & 16 \\
5 & 0.997 & 25 & 17 & 0.996 & 16 \\
7 & 0.997 & 22 & \textbf{19} & \textbf{0.997} & \textbf{15} \\
9 & 0.996 & 19 & \textbf{21} & \textbf{0.997} & \textbf{15} \\
11 & 0.996 & 18 & \textbf{23} & \textbf{0.996} & \textbf{16} \\
13 & 0.996 & 16 & \textbf{25} & \textbf{0.995} & \textbf{15} \\ \bottomrule
\end{tabular}}
\end{table}

\begin{figure}[h]
\centering
\subfloat[$l = n/N$]{\includegraphics[width=8cm]{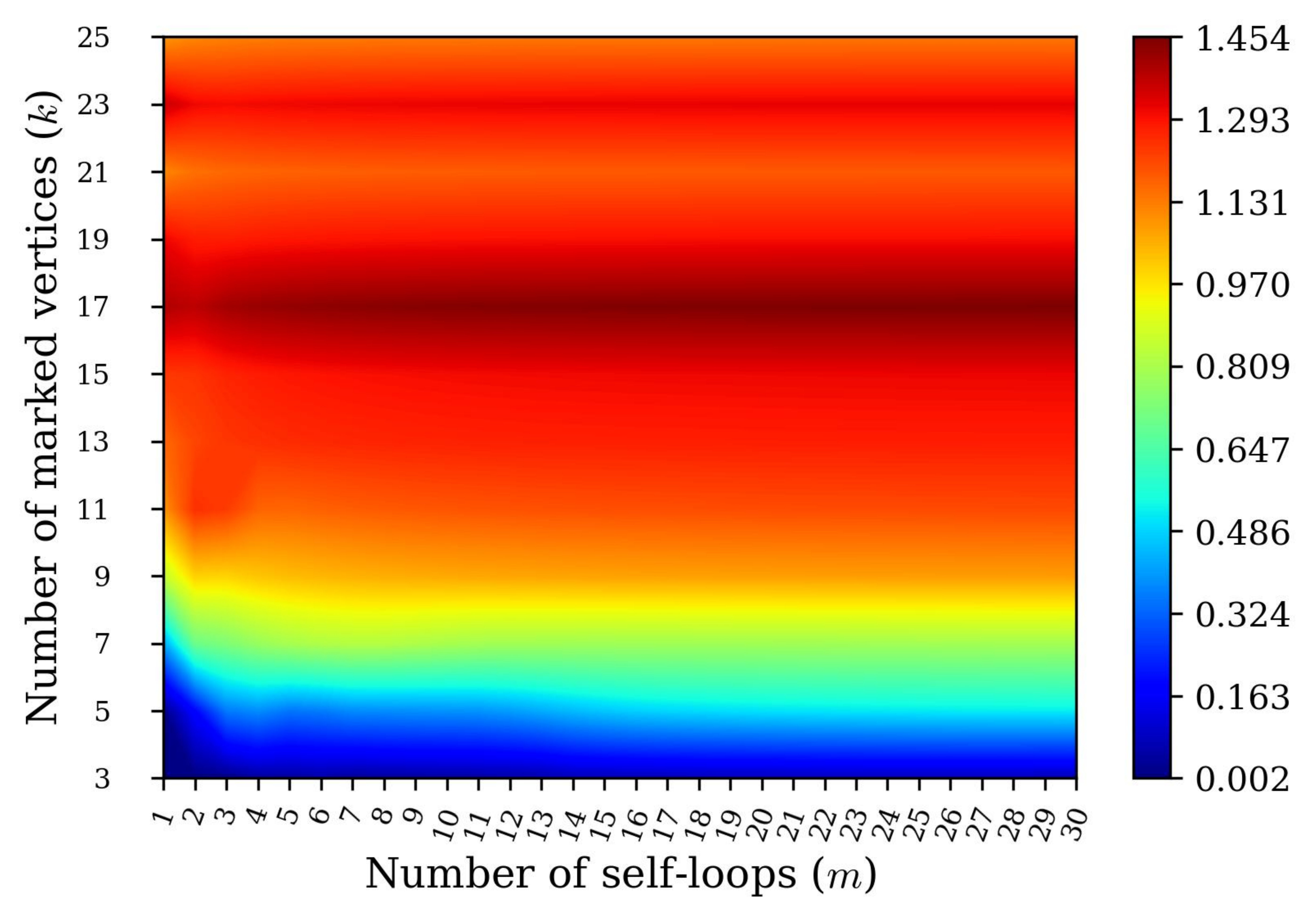}
\label{fig:probability-distribution-neighbors-non-neighbors-a-standard-deviation}}
\subfloat[$l = (n/N)\cdot k$]{\includegraphics[width=8cm]{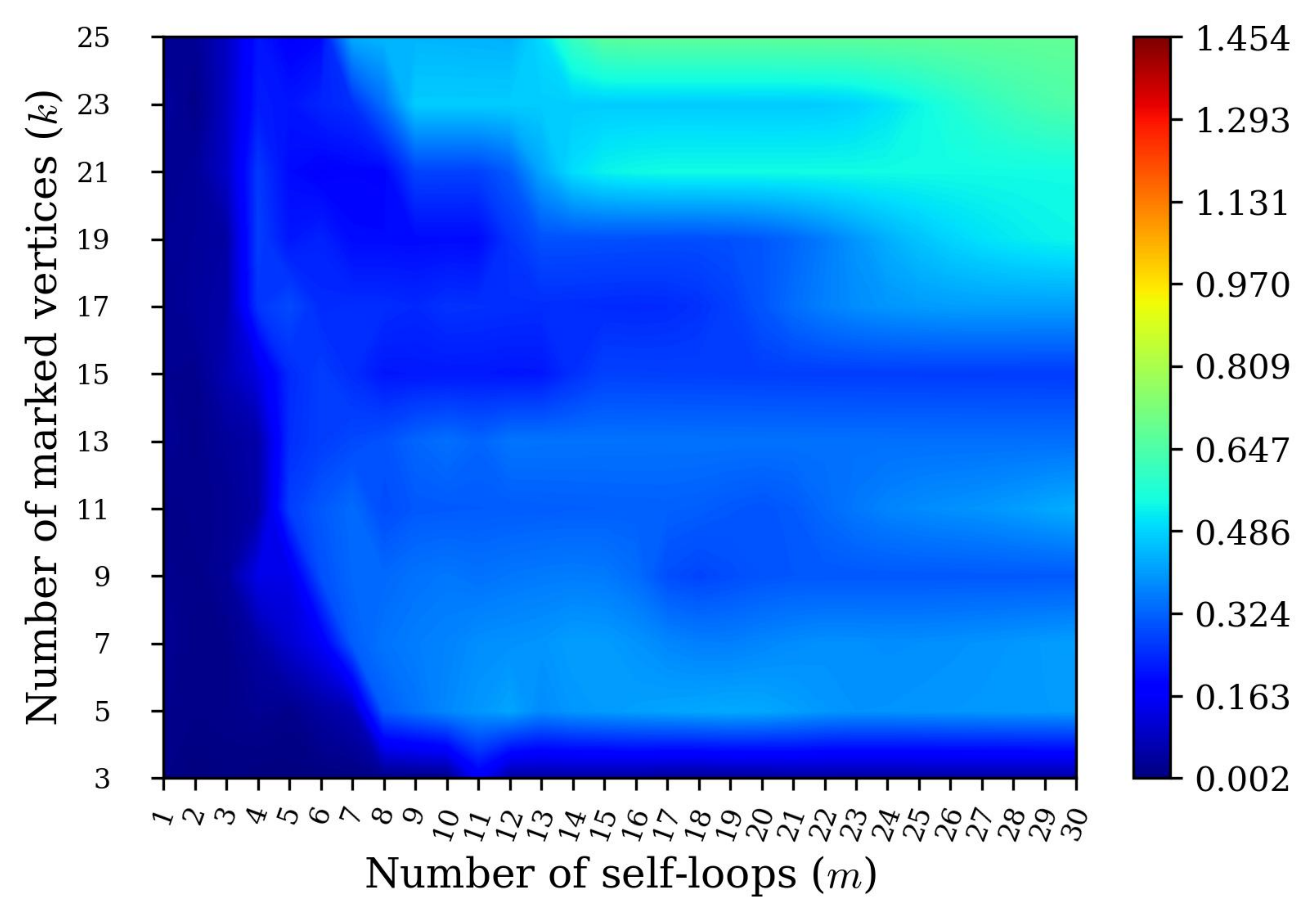}\label{fig:probability-distribution-neighbors-non-neighbors-b-standard-deviation}}\\
\subfloat[$l = n^{2}/N$]{\includegraphics[width=8cm]{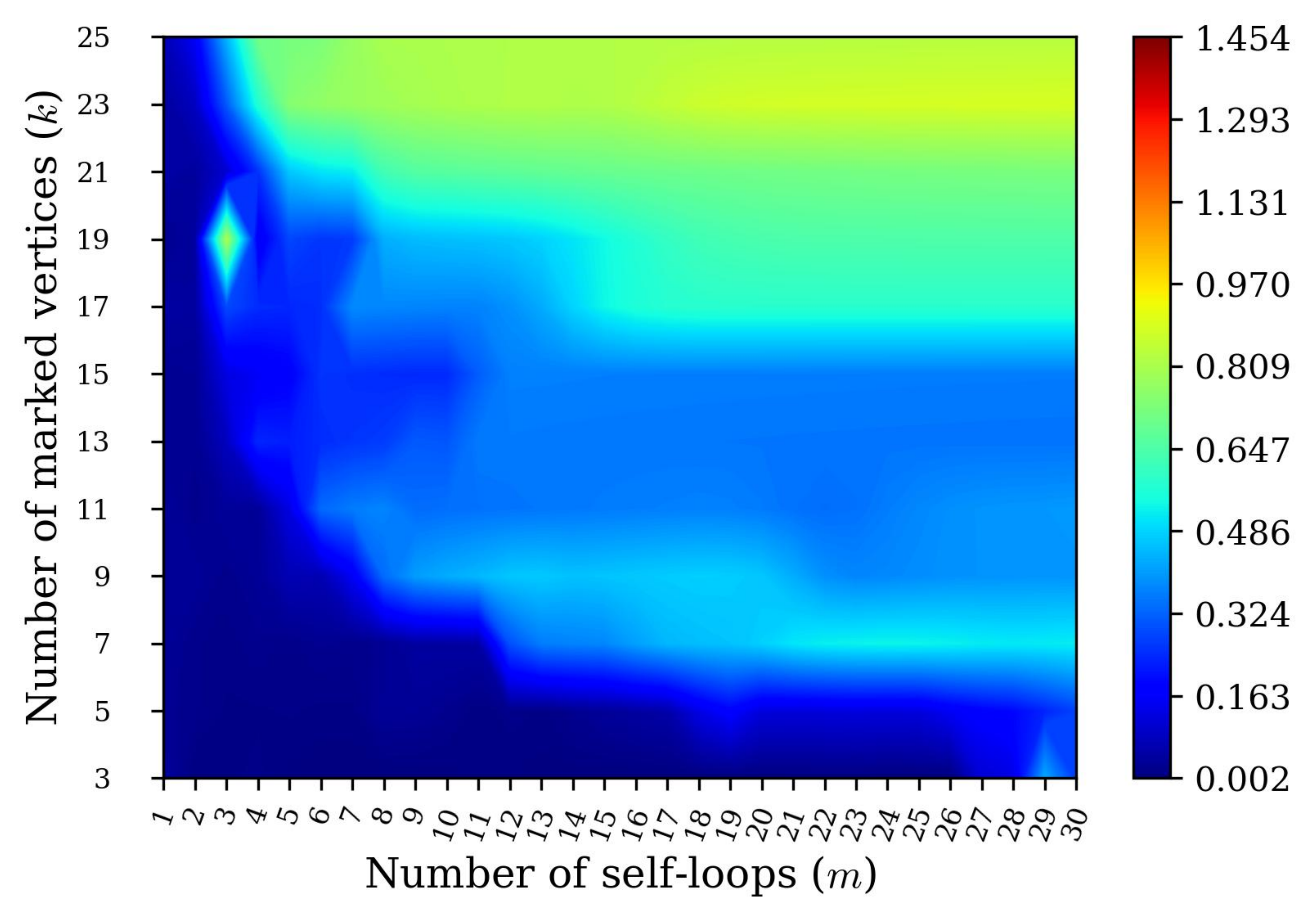}\label{fig:probability-distribution-neighbors-non-neighbors-c-standard-deviation}}
\subfloat[$l = (n^{2}/N)\cdot k$]{\includegraphics[width=8cm]{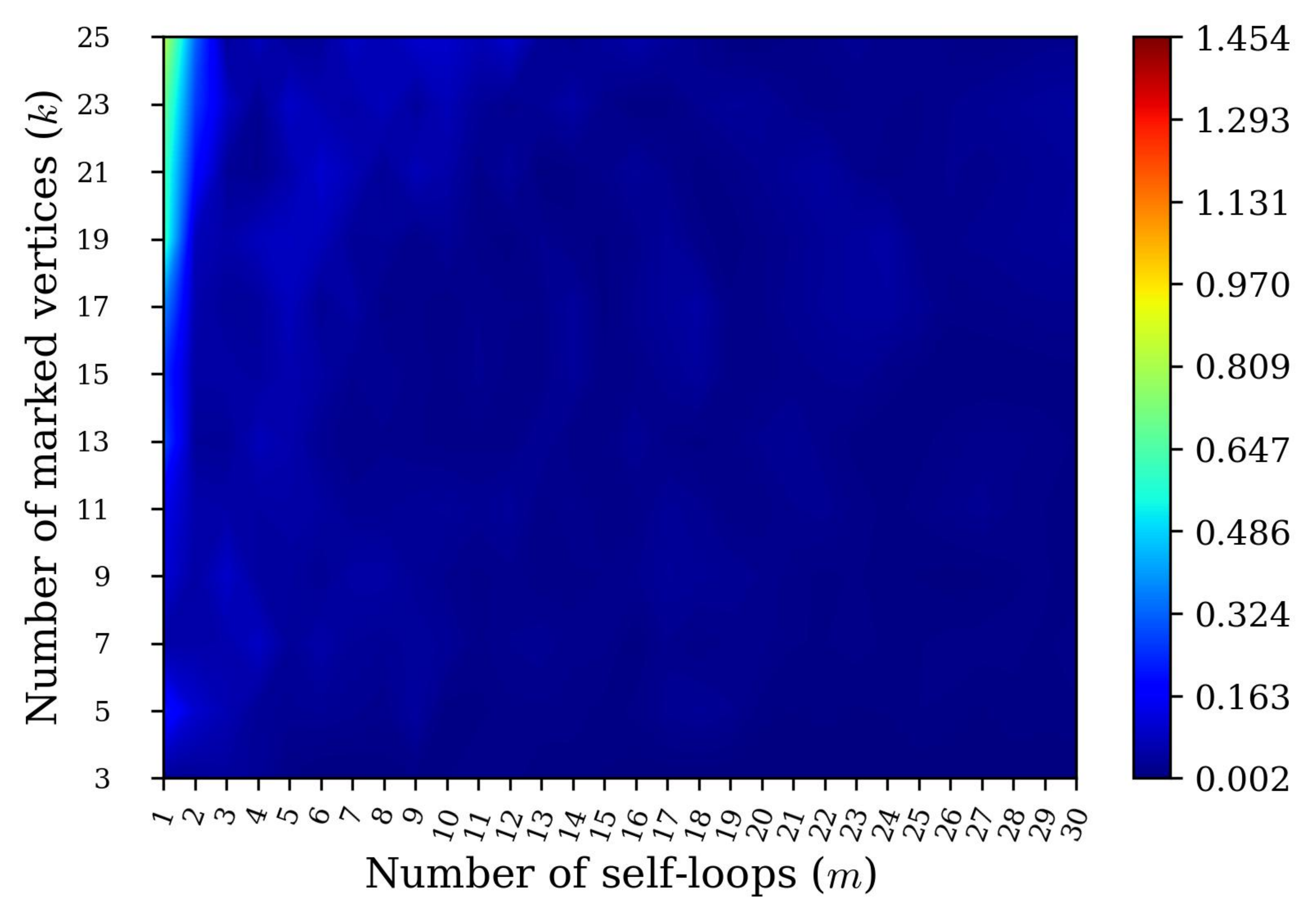}\label{fig:probability-distribution-neighbors-non-neighbors-d-standard-deviation}}
\caption{The coefficient of variation of the MSLQW-PPI to search for adjacent and non-adjacent marked vertices. The results are represented in percentage terms. \protect\subref{fig:probability-distribution-neighbors-non-neighbors-a-standard-deviation}, \protect\subref{fig:probability-distribution-neighbors-non-neighbors-b-standard-deviation}, \protect\subref{fig:probability-distribution-neighbors-non-neighbors-c-standard-deviation}, and \protect\subref{fig:probability-distribution-neighbors-non-neighbors-d-standard-deviation} represents the coefficient of variation of the results presented in Figures~\ref{fig:probability-distribution-neighbors-non-neighbors-a}, \ref{fig:probability-distribution-neighbors-non-neighbors-b}, \ref{fig:probability-distribution-neighbors-non-neighbors-c}, and \ref{fig:probability-distribution-neighbors-non-neighbors-d} for the weight values $l = n/N$, $l = (n/N) \cdot k$, $l = n^{2}/N$, and $l = (n^{2}/N)\cdot k$, respectively.}
\label{fig:coefficient-variation-neighbors-non-neighbors}
\end{figure}

As in \citet{desouza2023multiselfloop}, we analyzed whether the relative position of the non-adjacent marked vertices influences the results of the maximum probability of success. We also used the coefficient of variation to analyze the level of dispersion of the results. The relative position of the non-adjacent marked vertices also did not show significant influence considering a numerical precision of four digits. Fig.~\ref{fig:coefficient-variation-neighbors-non-neighbors} shows the coefficient of variation for the results presented in Fig.~\ref{fig:probability-distribution-neighbors-non-neighbors}. Variations around the mean value are small, however, the behavior shown is stable. The maximum success probabilities close to $1$ coincide with these small variations. Considering the weight value of the self-loop, in general, the weight $l = (n^{2}/N)\cdot k$ indicated minor variability.

\subsection{Analyzing the search with adjacent marked vertices}
\label{sec:analyzing-the-search-with-adjacent-marked-vertices}

\begin{figure}[h]
\centering
\subfloat[$l = n/N$]{\includegraphics[width=8cm]{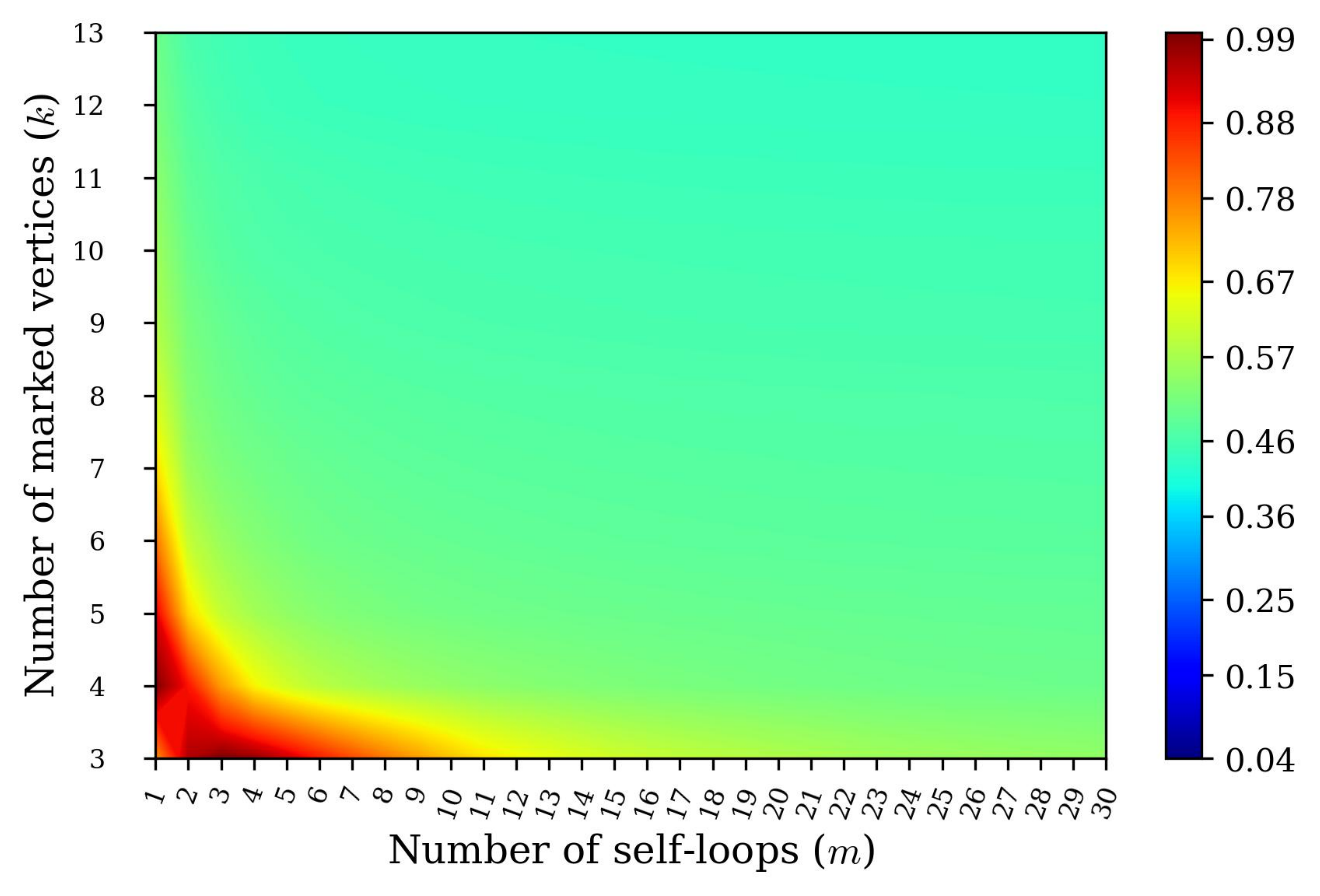}
\label{fig:probability-distribution-neighbors-a}}
\subfloat[$l = (n/N)\cdot k$]{\includegraphics[width=8cm]{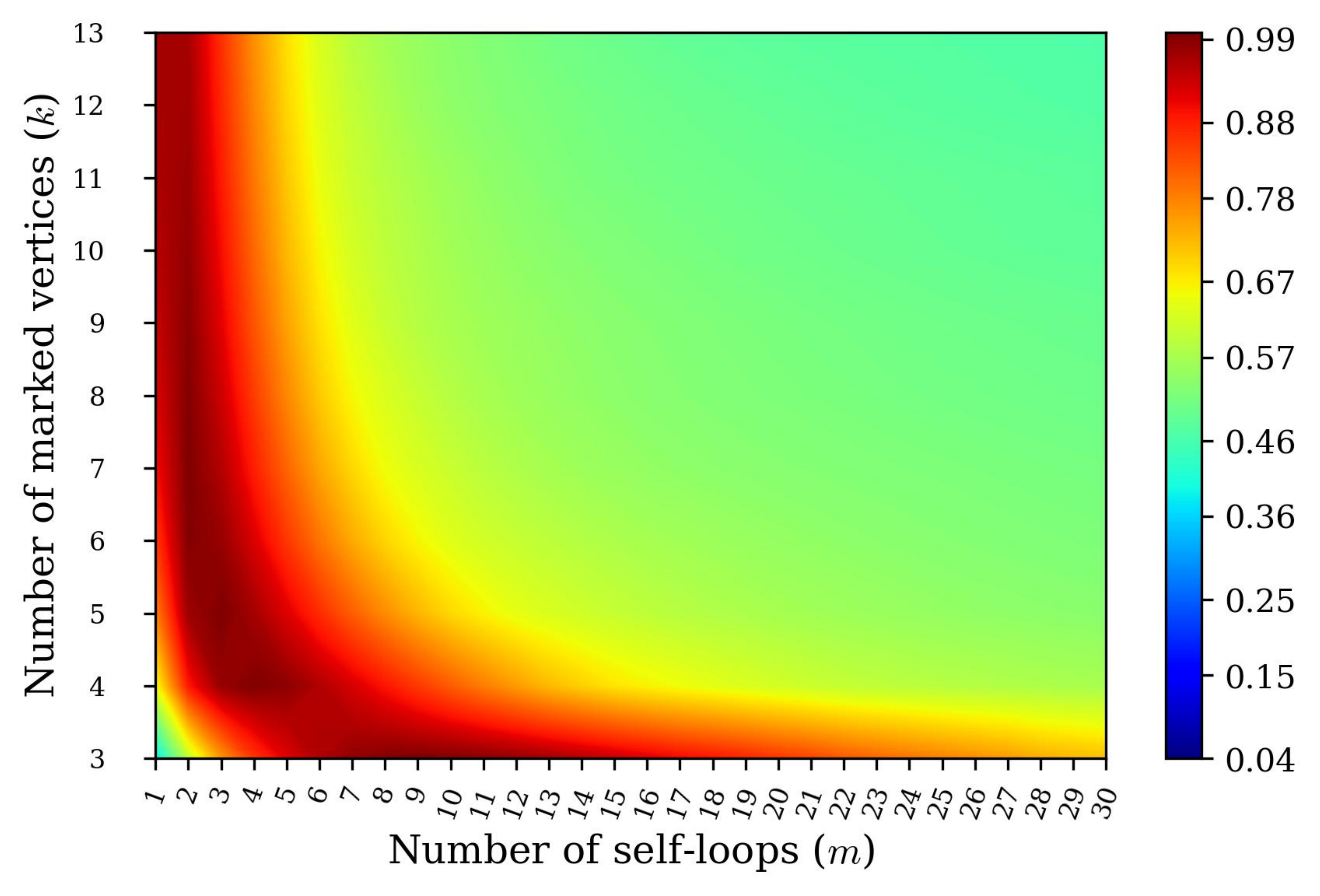}\label{fig:probability-distribution-neighbors-b}}\\
\subfloat[$l = n^{2}/N$]{\includegraphics[width=8cm]{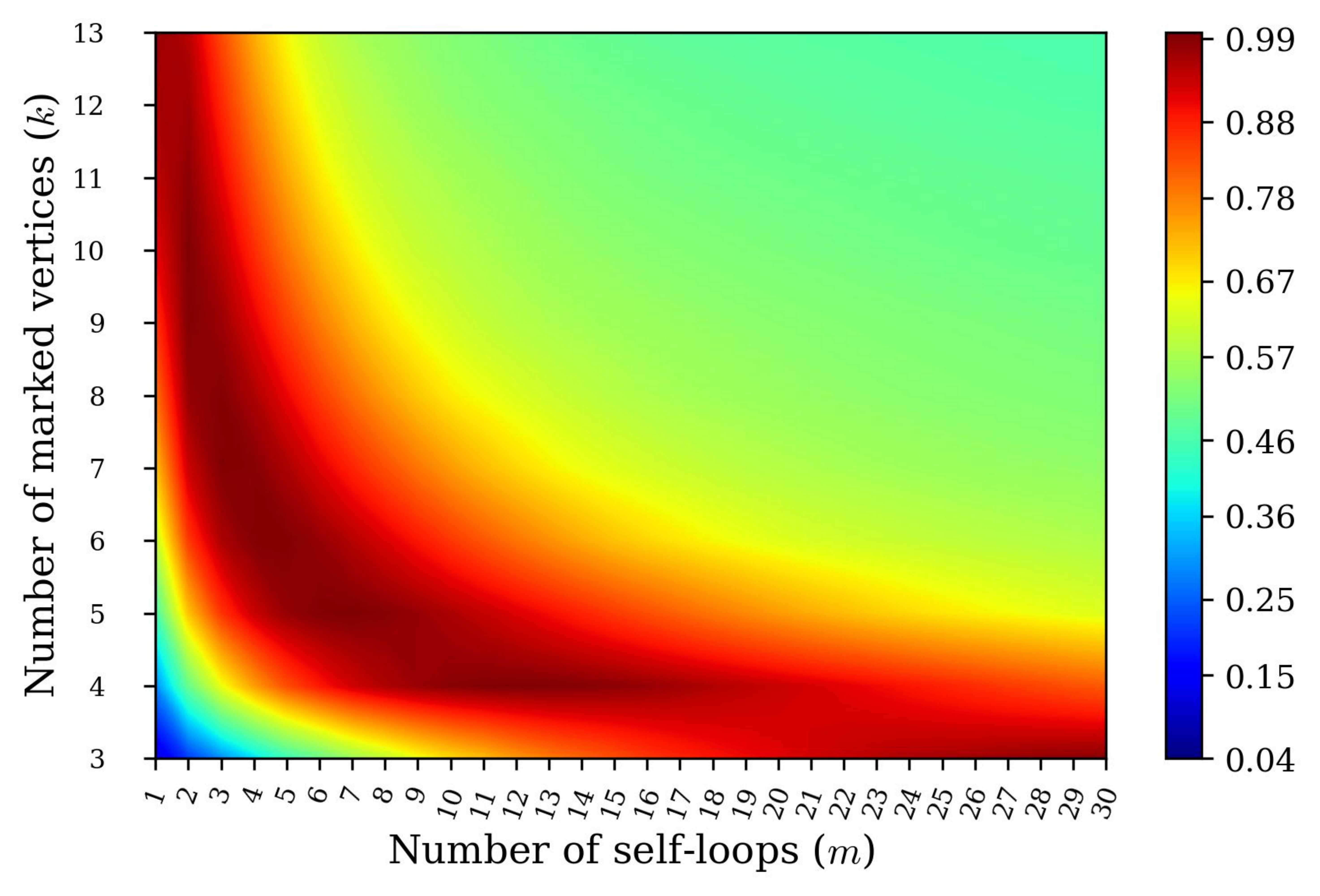}\label{fig:probability-distribution-neighbors-c}}
\subfloat[$l = (n^{2}/N)\cdot k$]{\includegraphics[width=8cm]{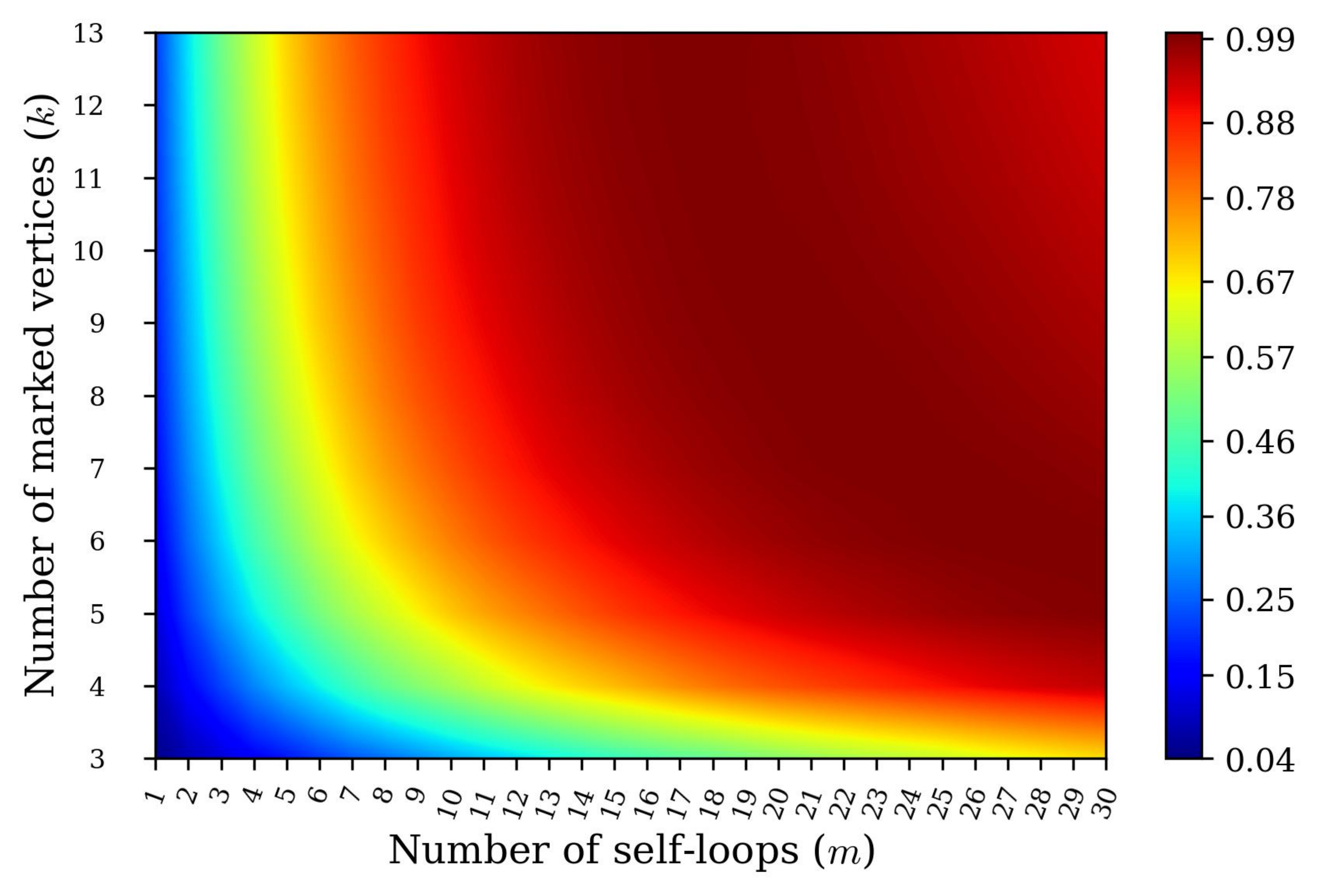}\label{fig:probability-distribution-neighbors-d}}
\caption{The probability of success of the MSLQW-PPI to search for adjacent marked vertices with $n = 12$ and $N = 4096$ vertices. \protect\subref{fig:probability-distribution-neighbors-a} weight value $l = n/N$. \protect\subref{fig:probability-distribution-neighbors-b} weight value $l = (n/N) \cdot k$. \protect\subref{fig:probability-distribution-neighbors-c} weight value $l = n^{2}/N$. \protect\subref{fig:probability-distribution-neighbors-d} weight value $l = (n^{2}/N)\cdot k$.}
\label{fig:probability-distribution-neighbors}
\end{figure}

The simulations performed in the samples of the previous scenario were necessary so that we could obtain the average behavior based on the relative position of the non-adjacent marked vertices. In this scenario, let us analyze only adjacent vertices. According to \citet{nahimovs2019adjacent}, when there are two adjacent marked vertices, a stationary state occurs. In this case, the maximum probability of success obtained in our simulations was approximately $p = 0.02$ for all weights. Now, consider $k \geqslant 3$. Fig.~\ref{fig:probability-distribution-neighbors-a} shows the probability of success for the weight $l = n/N$. Comparing the results presented by \citet{souza2021lackadaisical} and \citet{desouza2023multiselfloop} to search for multiple marked vertices and a single self-loop we had an improvement in the success probability for $k = 3$ marked vertices which were $p = 0.745$ and evolved to $p = 0.999$ with $m = 3$ self-loops. 

Fig.~\ref{fig:probability-distribution-neighbors-b} shows the probability of success for the weight $l = (n/N)\cdot k$. Compared to the results obtained by \citet{souza2021lackadaisical} for searching multiple adjacent marked vertices using a single self-loop, there was an improvement in the probability of success. As we can see in Table~\ref{tab:weight-and-self-loops-adjacent-n-N-K}, two results are significant, the search for $k = 3$ marked vertices, with $9$ self-loops allowed to increase the probability from $p = 0.386$ to $p = 0.999$. For $k = 4$ marked vertices, with $4$ self-loops allowed to increase the probability from $p = 0.639$ to $p = 0.996$. Fig.~\ref{fig:probability-distribution-neighbors-non-neighbors-b} shows the results where both adjacent and non-adjacent vertices are marked. Analyzing the cases where the number of marked vertices is the same, i.e., $k = \{3, 5, 7, 9, 11\}$ with $m = 2$ self-loops, although the scenarios are different, the maximum probability of success remained above $p = 0.99$.

Again, let us analyze two different scenarios for the same weight values. Fig.~\ref{fig:probability-distribution-neighbors-c} shows the success probabilities for searching only adjacent marked vertices while Fig.~\ref{fig:probability-distribution-neighbors-non-neighbors-c} shows the success probabilities for searching adjacent and non-adjacent marked vertices both using the weight $l = n^{2}/N$. Note that the behaviors are very similar, however, in the scenario where there are only adjacent marked vertices, which is the case in Fig.~\ref{fig:probability-distribution-neighbors-c}, a greater number of self-loops is necessary when the density of adjacent marked vertices is small. Table~\ref{tab:comp-weight-and-self-loops-adjacent-adjacent-4c-5c} shows the comparison between the success probabilities and the number of self-loops for the cases where the number of marked vertices is the same. Again, note that the results are similar except for $k = 3$, where there was a significant increase in the number of self-loops.

\begin{table}[h]
\setlength{\tabcolsep}{10pt}
\centering
\caption{Comparison between the success probabilities and the number of self-loops to search for adjacent marked vertices with the weight $l = (n/N)\cdot k$. (\ref{tab:weight-and-self-loops-adjacent-and-non-adjacent-n-N-k-previous}) shows the results obtained by \citet{souza2021lackadaisical} using a single self-loop. (\ref{tab:weight-and-self-loops-adjacent-n-N-k-current}) shows the results in this work using multiple self-loops.}
\label{tab:weight-and-self-loops-adjacent-n-N-K}
\subfloat[]{
\label{tab:weight-and-self-loops-adjacent-and-non-adjacent-n-N-k-previous}
\begin{tabular}{cc}
\toprule
$k$ & $p$ \\ \midrule
\textbf{3} & \textbf{0.386} \\
\textbf{4} & \textbf{0.639} \\
5 & 0.783 \\
6 & 0.853 \\
7 & 0.889 \\
8 & 0.916 \\
9 & 0.937 \\
10 & 0.942 \\
11 & 0.945 \\ \bottomrule
\end{tabular}}
\hspace{0.5cm}
\centering
\subfloat[]{
\label{tab:weight-and-self-loops-adjacent-n-N-k-current}
\begin{tabular}{ccc}
\toprule
$k$ & $p$ & $m$ \\ \midrule
\textbf{3} & \textbf{0.999} & \textbf{9} \\
\textbf{4} & \textbf{0.996} & \textbf{4} \\
5 & 0.997 & 3 \\
6 & 0.994 & 2 \\
7 & 0.997 & 2 \\
8 & 0.994 & 2 \\
9 & 0.990 & 2 \\
10 & 0.982 & 2 \\
11 & 0.975 & 2 \\ \bottomrule
\end{tabular}}
\end{table}

Now, consider Fig.~\ref{fig:probability-distribution-neighbors-d}. It shows the success probabilities to search for adjacent marked vertices using the self-loop weight $l = (n^{2}/N) \cdot k$. Compared with the results of the scenario presented in Fig. \ref{fig:probability-distribution-neighbors-non-neighbors-d} we notice a very similar behavior where for a small density of marked vertices a greater number of self-loops is necessary, however, when this density of marked vertices increases, the number of self-loops decreases to the point of approaching the results presented by \citet{desouza2023multiselfloop} for the same weight value $l = (n^{2}/N)\cdot k$. Table~\ref{tab:weight-and-self-loops-adjacent-5d} shows the number of self-loops needed to obtain the maximum probabilities of success. Comparing with the results presented in Table~\ref{tab:weight-and-self-loops-adjacent-and-non-adjacent-n-2-N-k} for the same numbers of marked vertices, it is possible to see that, a greater number of self-loops are required to achieve success probabilities close to $1$ when there are only adjacent marked vertices. However, for $k = 3$, $m = 30$ was insufficient.

\begin{figure}[h]
\centering
\subfloat[Adjacents.]{\includegraphics[width=8cm]{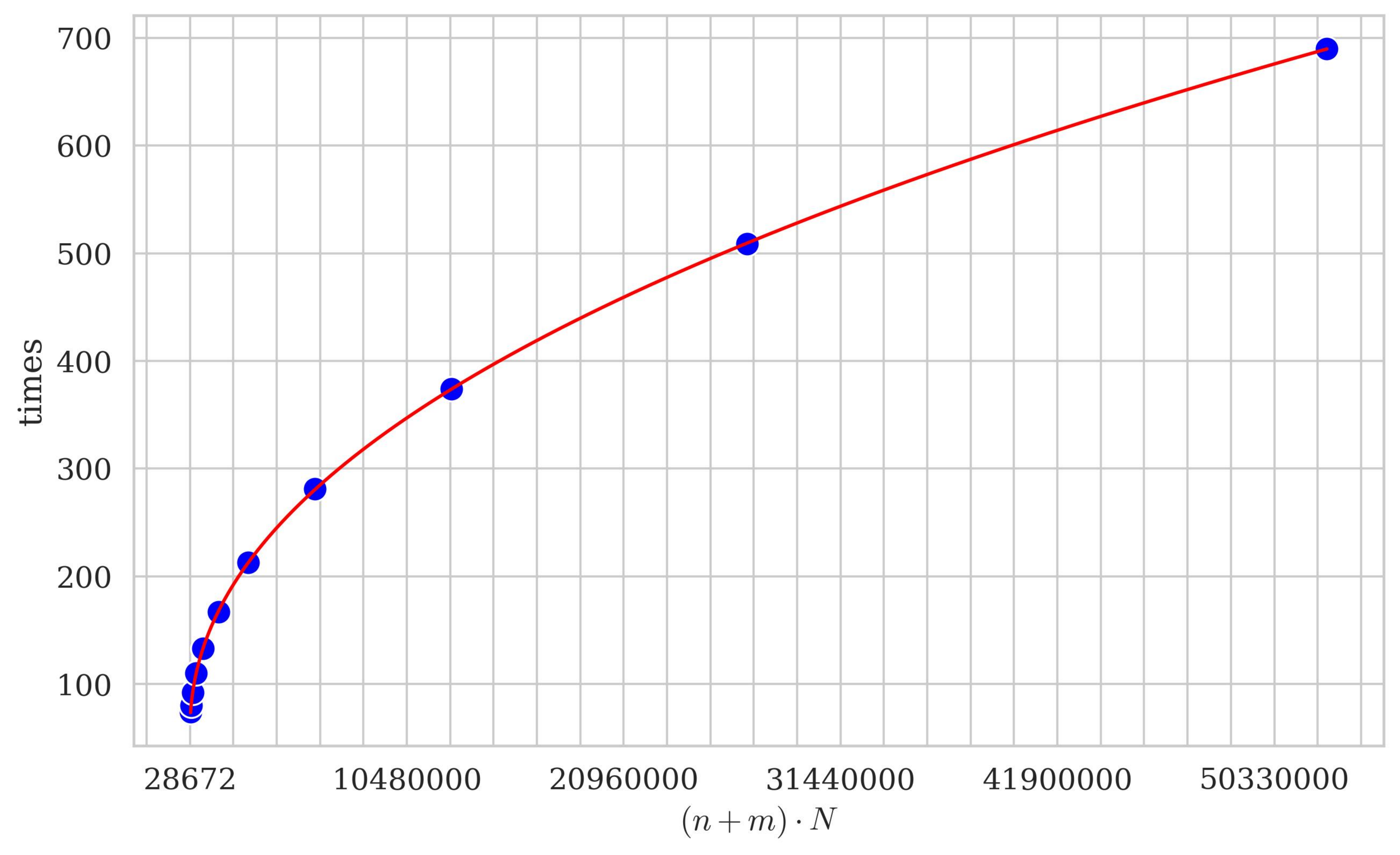}
\label{fig:hypercube-curve-fitting-one-self-loops-non-neighbors-n-2-N-k}}
\subfloat[Adjacents and non adjacents.]{\includegraphics[width=8cm]{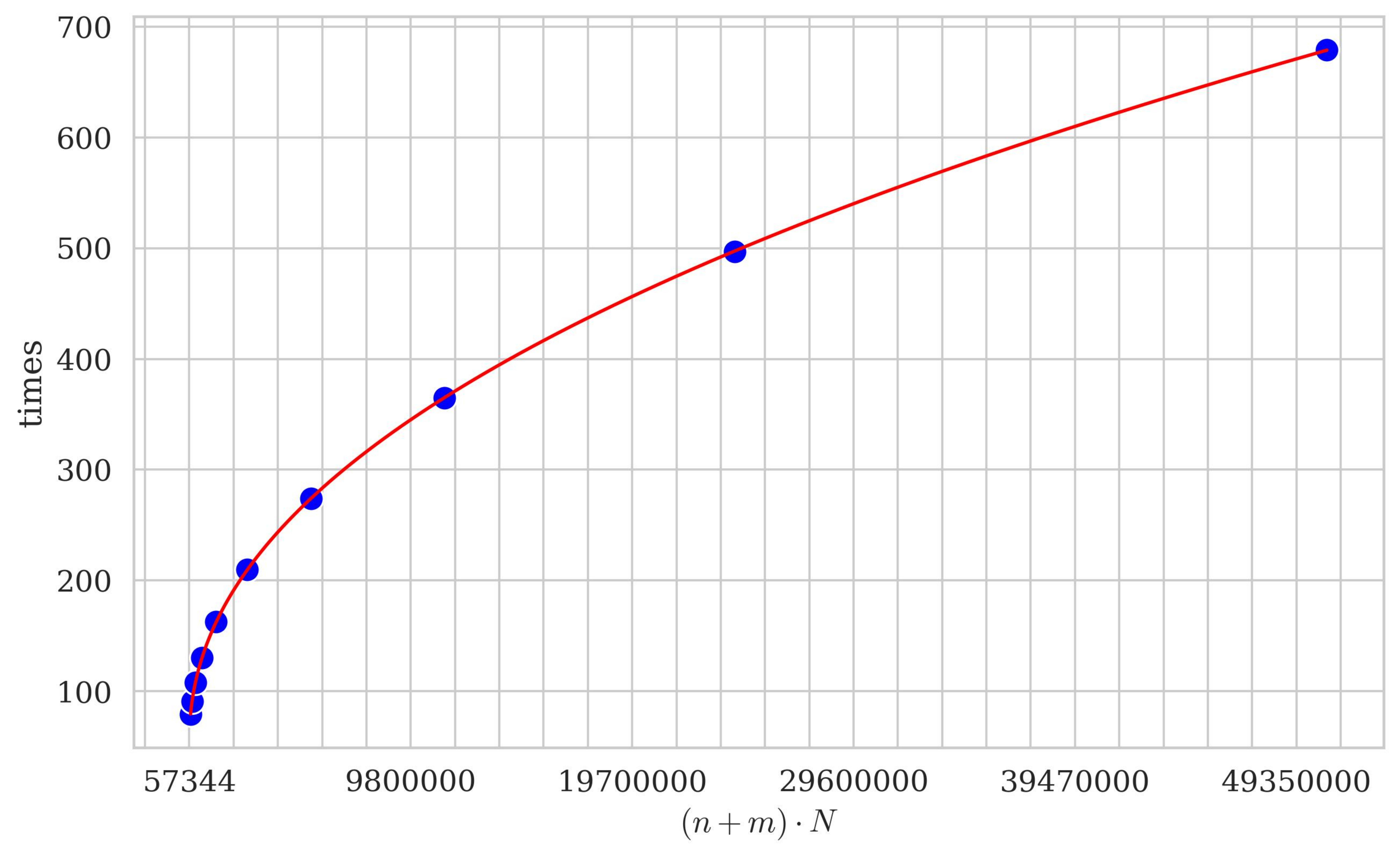}
\label{fig:hypercube-curve-fitting-multiple-self-loops-non-neighbors-n-2-N-k}}
\caption{The time complexity of the algorithm relative to the size of the hypercube. The solid red line represents the estimated curve and the blue dots are the numerical simulation values of the quantum walk.}
\label{fig:hypercube-curve-fitting-one-multiple-self-loops-non-neighbors-n-2-N-k}
\end{figure}

\begin{figure*}[]
    \centering
    \subfloat[Adjacents]{\includegraphics[width=8cm]{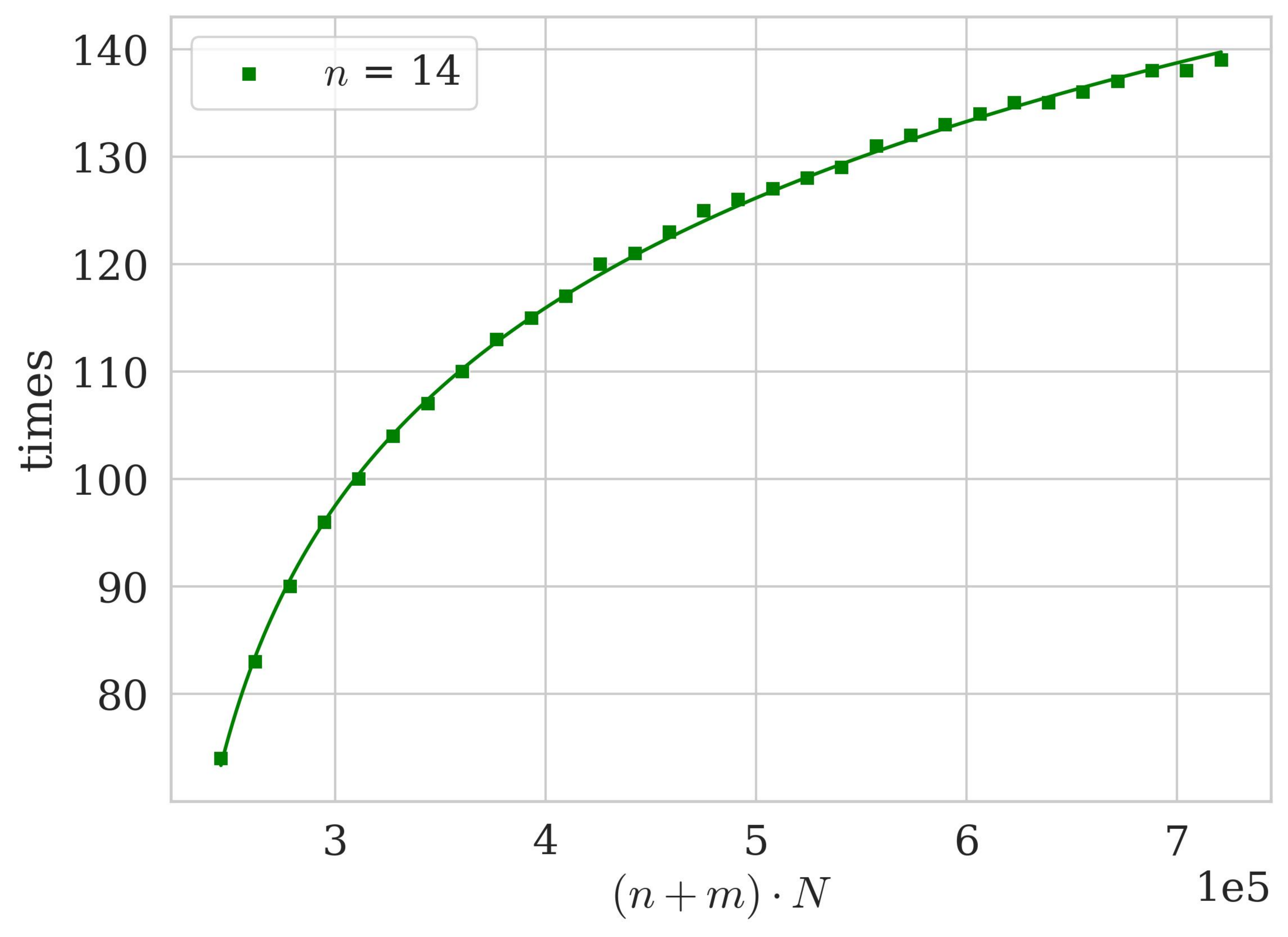}
    \label{fig:hypercube_curve_fitting_self-loops_n_12}}
    \subfloat[Adjacents and non adjacents]{\includegraphics[width=8cm]{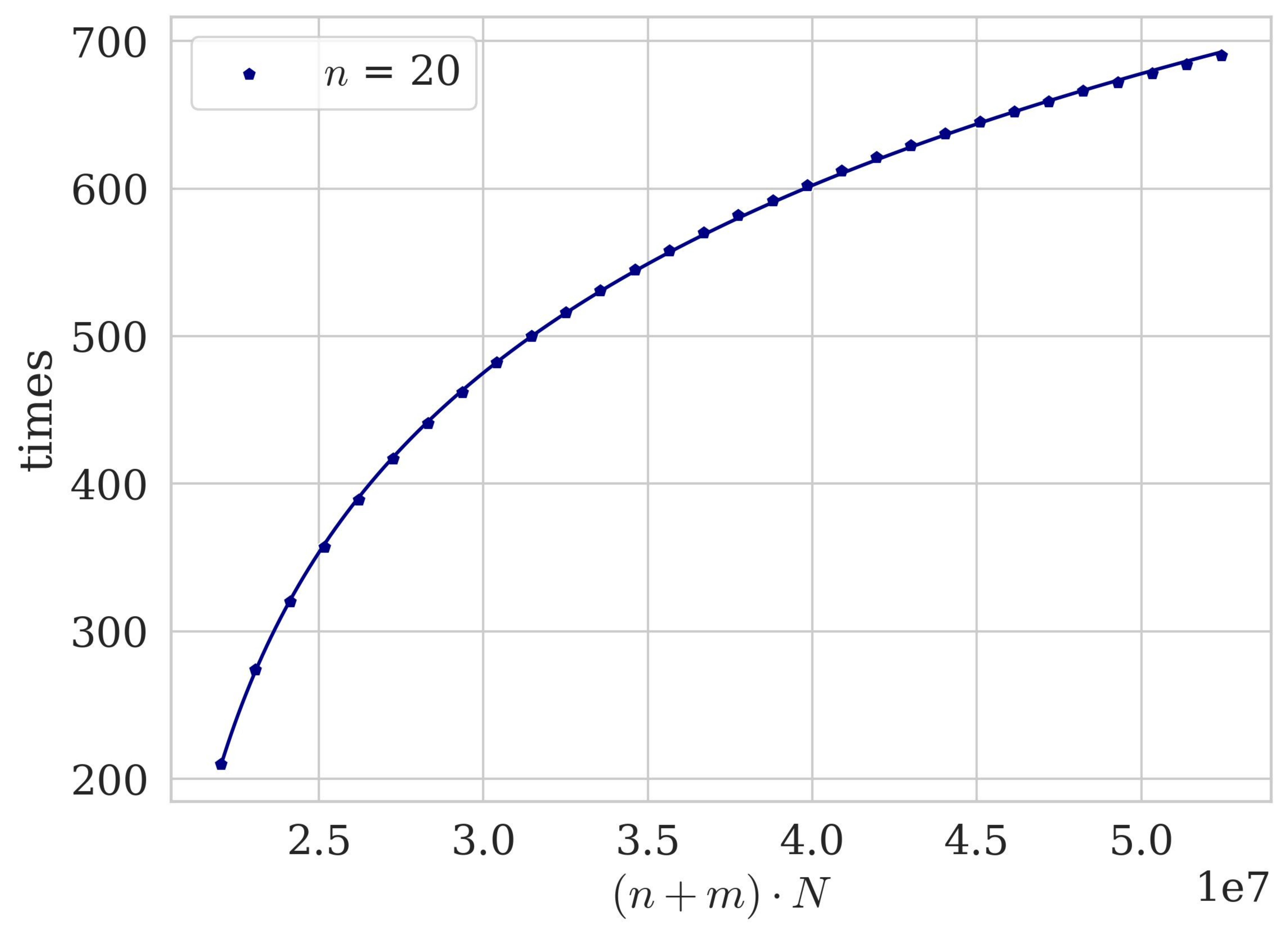}
    \label{hypercube_curve_fitting_self-loops_n_13}}\\
    \subfloat[Adjacents]{\includegraphics[width=8cm]{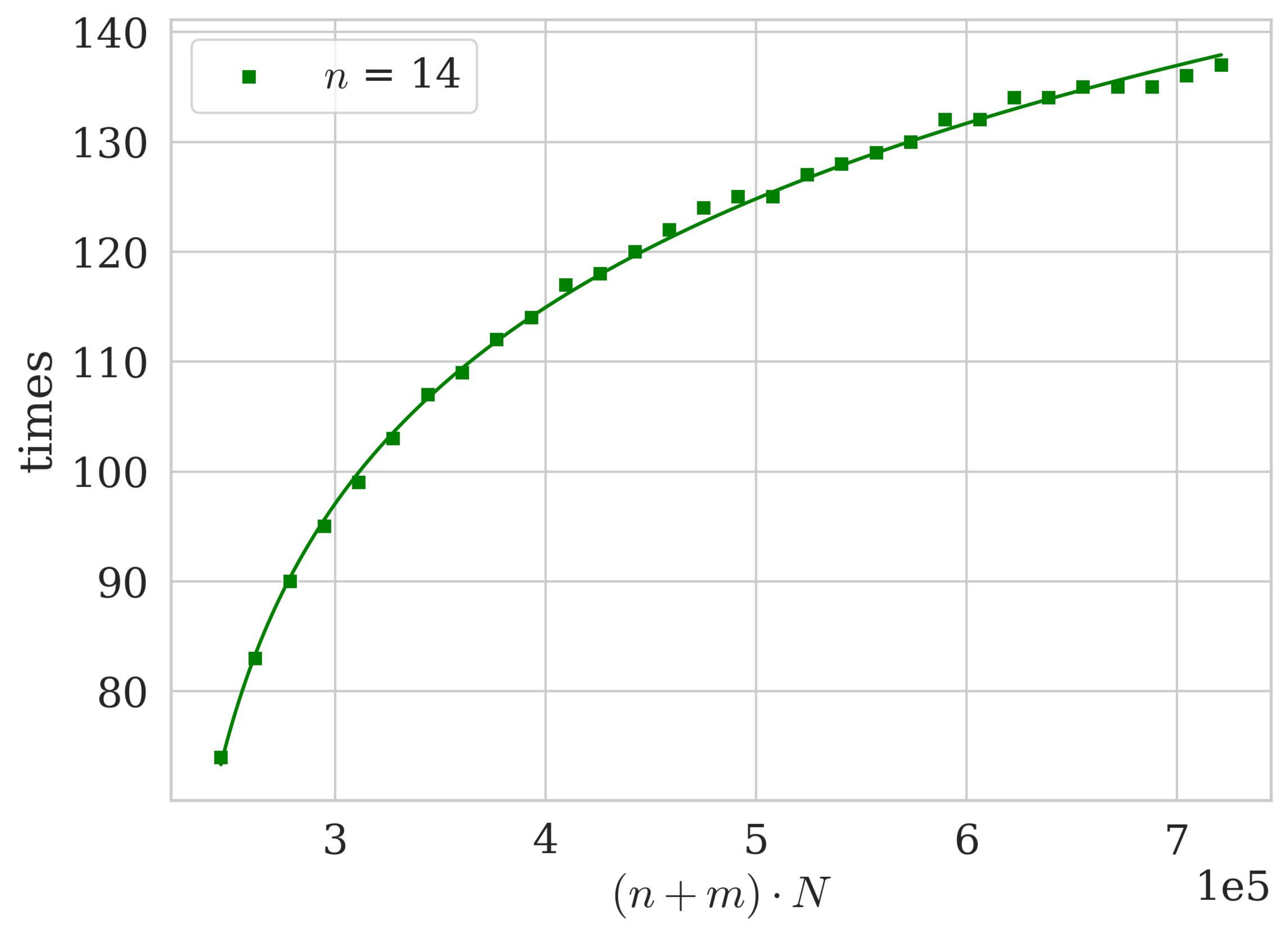}
    \label{fig:hypercube_curve_fitting_self-loops_n_14}}
    \subfloat[Adjacents and non adjacents]{\includegraphics[width=8cm]{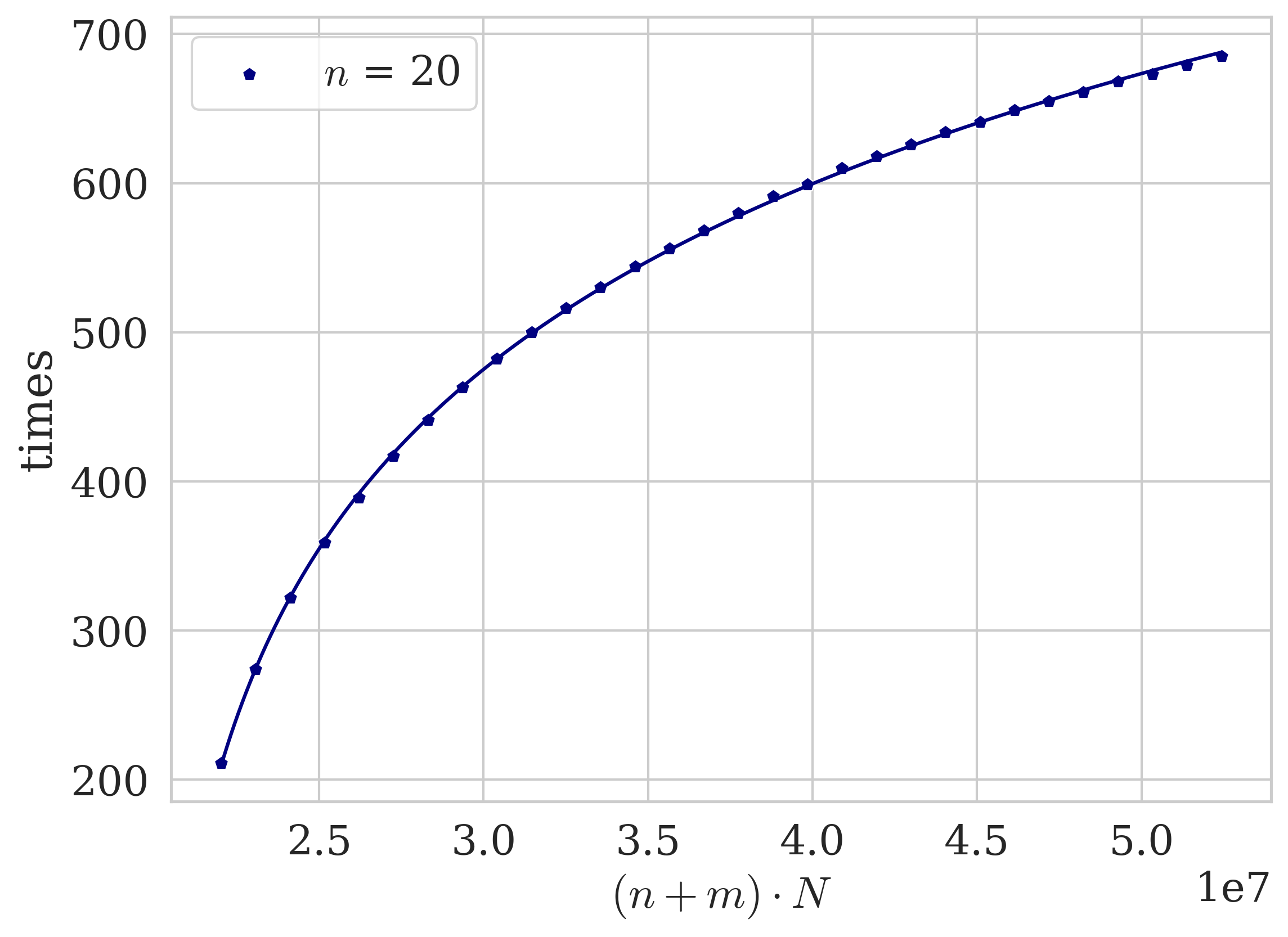}
    \label{fig:hypercube_curve_fitting_self-loops_n_15}}
    \caption{The algorithm's time complexity concerning the number of self-loops at each vertex. The solid lines represent the estimated curves. The points are the values from numerically simulating the quantum walk. For each figure, $n$ is a constant.}
    \label{fig:hypercube-curve-fitting-self-loops-non-neighbors-n-2-N-k}
\end{figure*}

To obtain the complexity of the algorithm proposed by \citet{desouza2023multiselfloop}. applied to the search for adjacent vertices, two analyses were performed. The first analysis described in Fig.~\ref{fig:hypercube-curve-fitting-one-multiple-self-loops-non-neighbors-n-2-N-k} shows the runtime complexity when $N = 2^{n}$ is changed. The second analysis described in Fig.~\ref{fig:hypercube-curve-fitting-self-loops-non-neighbors-n-2-N-k} shows how the runtime complexity behaves when $m$ self-loops are added at each vertex of the hypercube. As in \citet{desouza2023multiselfloop}, the computational cost when we consider the hypercube size is $O(\sqrt{((n + m) })$. The computational cost where $m$ self-loops varies is $O(\log{(m)})$.

\begin{table}[]
\centering
\caption{Comparison between the probability of success and number of self-loops for two different scenarios. The column for \ref{fig:probability-distribution-neighbors-non-neighbors-c} represents the results of searching for adjacent and non-adjacent vertices and the column for \ref{fig:probability-distribution-neighbors-c} to search for adjacent vertices.}
\label{tab:comp-weight-and-self-loops-adjacent-adjacent-4c-5c}
\begin{tabular}{lcccl}
\toprule
 & \multicolumn{4}{c}{Figures} \\ \cmidrule{2-5} 
 & \multicolumn{2}{c}{\ref{fig:probability-distribution-neighbors-non-neighbors-c}} & \multicolumn{2}{c}{\ref{fig:probability-distribution-neighbors-c}} \\ \cmidrule(lr){2-3}\cmidrule(lr){4-5} 
k & p & m & p & m \\ \midrule
3 & 0.999 & 12 & 0.991 & 30 \\
5 & 0.997 & 5 & 0.998 & 7 \\
7 & 0.996 & 3 & 0.995 & 3 \\
9 & 0.996 & 2 & 0.995 & 2 \\
11 & 0.983 & 2 & 0.986 & 2 \\ \bottomrule
\end{tabular}
\end{table}

\begin{table}
\centering
\setlength{\tabcolsep}{15pt}
\caption{Maximum success probability and number of self-loops to search for adjacent marked vertices with the weight $l = (n^{2}/N)\cdot k$.}
\label{tab:weight-and-self-loops-adjacent-5d}
\centering
\label{tab:weight-and-self-loops-adjacent-n-2-N-k}
\begin{tabular}{ccc}
\toprule
$k$ & $p$ & $m$ \\ \midrule
3 & \textbf{0.681} & \textbf{30} \\
4 & \textbf{0.943} & \textbf{30} \\
5 & 0.995 & 30 \\
6 & 0.999 & 28 \\
7 & 0.998 & 24 \\
8 & 0.998 & 21 \\
9 & 0.998 & 20 \\
10 & 0.996 & 19 \\
11 & 0.997 & 18 \\
12 & 0.997 & 18 \\
13 & 0.997 & 17 \\ \bottomrule
\end{tabular}
\end{table}


\section{Conclusions}
\label{sec:conclusions}

In this work, we analyzed the application of MSLQW-PPI in two scenarios based on the type of marked vertices: adjacent and non-adjacent. In the first scenario, the two types of marked vertices were searched. Here, we analyzed the relative position of the non-adjacent marked vertices to verify your influence on the results about adjacent vertices, for this, the coefficient of variation was also used to verify the dispersion of results around the maximum success probability mean value as \citet{desouza2023multiselfloop}. In the second scenario, we analyzed only the adjacent marked vertices. The dependence on the self-loop weight value is inherent to the lackadaisical quantum walk. Therefore, the composition of the weights is important. Thus, all analyses were made considering the four weight values. However, when applied to MSLQW-PPI a strategy of weight distribution is equally necessary because due to the use of the multiple self-loop. The strategy of weight distribution in this work was the same used by \citet{desouza2023multiselfloop}, i.e., the weight $l = l'/m$.

In the first scenario, to search for adjacent and non-adjacent vertices, according to the results, the relative position of non-adjacent marked vertices does not have a significant influence, considering a numerical precision of four digits. The results obtained by \citet{souza2021lackadaisical,desouza2023multiselfloop} to search for non-adjacent marked vertices indicate that the weight values influenced the maximum success probabilities according to weight values $l = (n/N)\cdot k$ and $l = (n^{2}/N)\cdot k$. Moreover, when we analyze the coefficient of variation, we saw that a minor variability coincides with the maximum probability of success. Fig.~\ref{fig:coefficient-variation-neighbors-non-neighbors} shows that the number of marked vertices influences the results causing a more significant variability. In the second scenario, the results obtained in the search for adjacent marked vertices present very similar results to the search for adjacent and non-adjacent marked vertices, except in some cases. For example, when the weight $l = (n^{2}/N)\cdot k$, the number of self-loops increases considerably. 

In summary, we conclude that for MSLQW-PPI, there is a dependence between the weight value, the vertex type, the number of marked vertices, and the number of self-loops needed to obtain success probabilities close to $1$. In the search for adjacent and non-adjacent marked vertices, the weight that presented the best results was $l = (n^{2}/N)\cdot k$. In the search for adjacent marked vertices, three of four weights presented the best results: $l = (n/N)\cdot k$, $l = (n^{2}/N)$, and $l = (n^{2}/N)\cdot k$. The results presented in Fig.~\ref{fig:probability-distribution-neighbors-d} show that from a certain $k$, the self-loops converge to a certain quantity. In future works, we intend to apply this methodology to evaluate the MSLQW-PPI in other $d$-regular structures with samples that contain adjacent marked vertices. We intend to verify the convergence of the number of multiple self-loops for a specific $m$ from a certain $k$ for the weight value $l = (n^{2}/N)\cdot k$ to search for adjacents marked vertices.

\section*{Acknowledgments}
\label{sec:acknowledgments}
Acknowledgments to the Science and Technology Support Foundation of Pernambuco (FACEPE) Brazil, Brazilian National Council for Scientific and Technological Development (CNPq), and Coordena\c{c}\~{a}o de Aperfei\c{c}oamento de Pessoal de N\'{i}vel Superior - Brasil (CAPES) - Finance Code 001 by their financial support to the development of this research.

\bibliographystyle{unsrtnat}
\bibliography{references}

\end{document}